\documentclass[12pt,english]{article}

\usepackage[utf8]{inputenc}
\usepackage[T1]{fontenc}

\usepackage{amsmath,amssymb,amsfonts}
\usepackage{bm,bbm}
\usepackage{booktabs}
\usepackage{color}
\usepackage[dvips,letterpaper]{geometry}
\usepackage{epsfig}
\usepackage{fullpage}
\usepackage{graphicx,psfrag,epsf}
\usepackage{indentfirst}
\usepackage{lineno}
\usepackage{natbib}
\usepackage{setspace}
\usepackage{subfigure}
\usepackage{times}
\usepackage{url}
\usepackage{verbatim}

\usepackage{multirow}
\usepackage[table,xcdraw]{xcolor}
\usepackage{adjustbox}
\usepackage{rotating}

\usepackage{caption}
\usepackage{adjustbox}

 \vspace {-0.7cm}

\title{\bf On a log-symmetric quantile tobit model applied to female labor supply data\footnote{This study was financed in part by the Coordenação de Aperfeiçoamento de Pessoal de Nível Superior - Brazil (CAPES) - Finance Code 001. Danúbia Rodrigues thanks CAPES/PROEX for the doctorate scholarship and José A. Divino and Helton Saulo acknowledge CNPq for financial support.}}

\author{
\normalsize
\textbf{Danúbia R. Cunha}$^{1}$\,, \textbf{Jose A. Divino}$^{1}$, \textbf{Helton Saulo}$^{2}$ \\[-0.05cm]
{\footnotesize $^{1}$Department of Economics, Catholic University of Brasilia, Brasilia, Brazil}\\[-0.05cm]
{\footnotesize $^{2}$Department of Statistics, University of Brasilia, Brasilia, Brazil}\\[-0.05cm]
}

\date{}

\begin{document}

%\linenumbers

\maketitle

\vspace {-0.7cm}
\noindent{\bf Abstract.} \emph{The classic censored regression model (tobit model) has been widely used in the economic literature. This model assumes normality for the error distribution and is not recommended for cases where positive skewness is present. Moreover, in regression analysis, it is well-known that a quantile regression approach allows us to study the influences of the explanatory variables on the dependent variable considering different quantiles. Therefore, we propose in this paper a quantile tobit regression model based on quantile-based log-symmetric distributions. The proposed methodology allows us to model data with positive skewness (which is not suitable for the classic tobit model), and to study the influence of the quantiles of interest, in addition to accommodating heteroscedasticity. The model parameters are estimated using the maximum likelihood method and an elaborate Monte Carlo study is performed to evaluate the performance of the estimates. Finally, the proposed methodology is illustrated using two female labor supply data sets. The results show that the proposed log-symmetric quantile tobit model has a better fit than the classic tobit model.}\\ 
\noindent{\bf Keywords:} Log-symmetric distributions; Quantile regression; Monte Carlo simulation; PSID, PNAD.\\
%\noindent{\bf JEL:} XXX; XXX.\\

% \vspace {-0.4cm}
% 
% \noindent{\bf Abstract.} \emph{Traduzir o resumo.
% }\\
% \noindent{\bf Keywords:} Distributions;;;;.\\
% \noindent{\bf JEL:} J20; C50.\\

\onehalfspacing

\section{Introduction}\label{sec:01}

In a sample, there is a possibility that the dependent variable cannot be observed in its entire domain or it is not observed for some of the individuals that incorporate it, namely, censoring is present in the data. This occurs when the information on the dependent variable is not fully available for some units of the sample. However, for these units, data on explanatory variables are known across their domain. Thus, in these cases, it is necessary to work with models that take censoring into account, such as the tobit model; see \cite{long:97}.

The classic tobit model was introduced by \cite{t:58} and has been widely used in economics, in addition to several other areas; see, for example, \cite{amemiya:84}, \cite{helsel:11} and \cite{bgls:17}. This model is used to left-censored dependent variables and was motivated by a study of the relationship between family spending on durable goods and family income. In this study, a portion of the dependent variable (family expenditure) was zero, that is, censored at a fixed limit value. Other studies in which the dependent variable is censored at zero for some observations include \cite{fair:77,fair:78} for modeling the number of extramarital affairs, \cite{jarque:87} for modeling family spending in various groups of commodities, and \cite{melenbergsoest:96} for modeling holiday expenses, among others.

The classic tobit model can also be used to estimate labor supply equations. Fixing working hours as the dependent variable, there is the possibility that the hours worked take the value zero. This happens when one or more individuals do not work, that is, for those who do not offer hours of work; see details in \cite{moffitt:82}. According to the labor supply model of \cite{heckmanmaCurdy:80}, which was discussed in 
\cite{islam:07}, the censored model is relevant in cases where the sample consists of individuals chosen at random and with hours reported as zero if the individual does not work. When it occurs, the techniques used for estimating linear models are inappropriate due to the presence of censoring.

One of the limitations of the classic tobit model \citep{t:58} is the assumption that the error term is normally
distributed. Although the normal distribution is widely used in several areas, it may not be appropriate to model dependent variables where positive skewness and/or light- and heavy-tails are present. In this context, the class of log-symmetric distributions represent important tools to overcome such problems. The log-symmetric distributions are a generalization of the log-normal distribution and have as its special cases distributions that have lighter or heavier tails than those of the log-normal, as well as bimodal distributions; see, for example, \cite{j:08}, \cite{vanegasp:15,vanegasp:16a,vanegasp:16b,vanegaspaula:17} and \cite{medeirosferrari:16}. In addition to the log-normal distribution, other examples of log-symmetric distributions are the log-Student-$t$, log-power-exponential and extended Birnbaum-Saunders, among others.

Although important, the use of an appropriate distribution to describe the error distribution of a tobit model does not provide a more comprehensive picture of the effect of the explanatory variables on the dependent variable. In this sense, quantile regression plays an important role, being able to model conditional quantiles as a function of explanatory variables; details on quantile regression can be seen in \cite{hn:07} and \cite{dfv:14}. In addition, quantile regression modeling is more efficient in cases where errors are not normally distributed or when the dependent variable has extreme value. 
%%\citep{RePEc:anp:en2001:102}.

In this context, the primary objective of this paper is to propose a quantile tobit model based on log-symmetric distributions. The strategy for the proposal of the new tobit model uses a reparameterization of the log-symmetric distributions proposed by \cite{ssls:20}, which has the quantile as one of its parameters. The secondary objectives are: (i) to obtain the maximum likelihood estimates of the model parameters; (ii) to carry out a Monte Carlo simulation to evaluate the performance of the maximum likelihood estimates; and (iii) to apply the proposed methodology to two real female labor supply data sets. The first application uses data from the
Panel Study of Income Dynamics (PSID) which is an American longitudinal household survey, whereas the second application uses data extracted from the Brazilian National Household Sample Survey (PNAD) carried out in 2015. The PSID data were studied by \cite{bgls:17} using the classical tobit model, and the PNAD data were filtered by the authors from the original data available at the Brazilian Institute of Geography and Statistics (IBGE) website\footnote{The use of the PSID and PNAD databases on female labor supply is relevant since the participation rate of women in the labor market has been a transforming factor in the labor market. In the United States case, for example, \cite{jacobsen99} claims that the increase in female participation in the labor market is the most striking economic statistic in the 20th century. As for Brazil, \cite{scomenfilho01} emphasize that there was a strong increase in female participation rates, especially for women with 1 to 11 years of study. On the other hand, \cite{barrosjatobamendoca95} highlight that the study of the mechanisms and motivations that explain the increase in the rate of female participation in the labor market, in addition to the fact that this rate is a basic socio-economic indicator, have boosted attention to the area. However, even today, women are less likely to participate in the labor market than men, in addition, they are also more likely to be unemployed in most countries of the world, according to a recent study by the International Labour Organization \citep{iol2018}. This justifies the use of the PSID and PNAD databases on female labor supply, given that the interest is in using censored data. In these data, part of the women do not work and for that reason they declare the income as being zero, which is classified as censored.}. The proposed approach can be seen as a generalization of the works of \cite{desousaetal:18} and \cite{slnb:20}. In general, the results of the applications showed that the proposed log-symmetric quantile tobit model provides better fit as compared to the classic tobit model.

The advantages of the proposed log-symmetric quantile tobit model over the classic normal tobit model are: (a) greater flexibility in terms of distributional assumption, since the log-symmetric class incorporates several special cases, such as the log-normal, log-Student-$t$, log-power-exponential, and extended Birnbaum-Saunders, among others; (b) greater flexibility for data modeling, allowing to consider the effects of explanatory variables on the dependent variable along the spectrum of the dependent variable, due to the quantile approach; and (c) flexibility to accommodate heteroscedasticity, since the proposed tobit model allows the insertion of explanatory variables in the dispersion parameter. On the other hand, the quantile approach proposed in this work differs from the existing models in the literature, since we introduce a quantile tobit model based on a reparametrization of the error distribution. The existing quantile tobit models studied in the literature, for example, \cite{powel:86}, \cite{buchinsky:98}, \cite{jilinzhang:12}, \cite{yuehong:12} and \cite{alhamzawiali:18}, are based on the minimization approach introduced by \cite{koenker:78,koenker:05}. Thus, the proposed approach is unprecedented in the literature for tobit models.

The rest of this paper proceeds as follows. In Section \ref{sec:2}, we briefly describe the class of log-symmetric distributions, both in its classical representation and in its reparameterization by quantile proposed by \cite{ssls:20}. In Section \ref{sec:3}, we introduce the log-symmetric quantile tobit model, and then provide details on estimation, interpretation of parameter estimates and residual analysis. 
In Section \ref{sec:4}, we carry out a Monte Carlo simulation to assess the performance of the maximum likelihood estimates. In Section \ref{sec:5}, two applications to PSID and PNAD data are studied. Finally, in Section \ref{sec:6}, we discuss conclusions and some possible future research on this topic.

\section{Log-symmetric distributions}\label{sec:2}

This section briefly describes the classical log-symmetric distributions \citep{vanegasp:16a} and those based on the quantile proposed by \cite{ssls:20}. The log-symmetric distributions reparameterized by the quantile, that is, those that have the quantile as one of their parameters, will be used to propose the log-symmetric quantile tobit model.

\subsection{Classical log-symmetric distributions}\label{sec:2.1}
A random variable $T$ follows a log-symmetric distribution with scale parameter $\lambda>0$ and power parameter $\phi>0$, if its probability density function and cumulative distribution function are given by 
\begin{equation}\label{eq:lg-pdf}
f_{T}(t;\lambda,\phi)
=
\dfrac{1}{\sqrt{\phi}\,t}\,
g\!\left(\dfrac{1}{\phi}\left[\log(t)-\log(\lambda)\right]^2\right),
\quad t>0,
\end{equation}
and
\begin{equation}\label{eq:lg-cdf}
 F_{T}(t;\lambda,\phi)=G\!\left(\dfrac{1}{\phi}\left[\log(t)-\log(\lambda)\right]^2 \right), \quad t>0,
\end{equation}
respectively, where $G(\omega)=\eta{\int^{\omega}_{-\infty} g(z^2)  \,\textrm{d}z }$, $\omega\in\mathbb{R}$, with $\eta$ being a normalizing constant and $g(\cdot)$ a density generator. In this case, the notation $T\sim\textrm{LS}(\lambda,\phi,g)$ is used. The $100q$-th quantile of $T\sim\textrm{LS}(\lambda,\phi,g)$ is given by
\begin{equation}\label{eq:lg-qf}
 Q=Q_{T}(q;\lambda,\phi)=\lambda\exp\big(\sqrt{\phi}\,G^{-1}(q)\big), \quad q\in(0,1),
\end{equation}
where $G^{-1}$ is the inverse of $G$ given in \eqref{eq:lg-cdf}. Table~\ref{tab:gfuncions} presents some density generators $g$ for some log-symmetric distributions; see details in \cite{vanegasp:16a}. Note that the generator $g$ may involve and extra parameter $\xi$.

\begin{table}[!ht] 
\centering
\small
\caption{\small{Density generator $g(u)$ for some log-symmetric distributions.}}\label{tab:gfuncions}
%%\vspace{0.15cm}
\begin{tabular}{llllll}
\hline Distribution                       &&             $g(u)$                                                              \\ %%  && $v(x)$ \\ %&& $\nu(z)$\\
\hline
Log-normal($\lambda,\phi$)                  && $\propto$ $\exp\left( -\frac{1}{2}u\right)$   
\\[1.5ex]%  &&  $1$\\[1.5ex] 
Log-Student-$t$($\lambda,\phi,\xi$)        &&  $\propto$ $\left(1+\frac{u}{\xi} \right)^{-\frac{\xi+1}{2}}$, $\xi>0$    
\\[1.5ex]  %&&  $\frac{\xi+1}{\xi+x^2}$\\[1.5ex] 
Log-power-exponential($\lambda,\phi,\xi$)     &&  $\propto$ $\exp\left( -\frac{1}{2}u^{\frac{1}{1+\xi}}\right)$, $-1<{\xi}\leq{1}$   

 \\[1.5ex]

Extended Birnbaum-Saunders($\lambda,\phi,\xi$)         &&  $\propto$ $\cosh(u^{1/2})\exp\left(-\frac{2}{\xi^2}\sinh^2(u^{1/2}) \right) $, $\xi>0$

\\\hline
\end{tabular}
\end{table}

\subsection{Quantile-based log-symmetric distributions}\label{sec:2.2}
Consider a fixed number $q \in (0,1)$ and $Q$ the $100q$-th quantile of $T\sim\textrm{LS}(\lambda,\phi,g)$ given in \eqref{eq:lg-qf}. Then, considering the one-to-one transformation $(\lambda,\phi) \mapsto (Q,\phi)$, \cite{ssls:20} 
proposed a reparameterization of the classical log-symmetric distribution, where the probability density function and the cumulative distribution function are given respectively by
\begin{equation}\label{eq:quant:ft}
f_{T}(t;Q,\phi)
=
\dfrac{1}{\sqrt{\phi}\,t}\,
g\!\left(\frac{1}{\phi} \left[ \log(t)-\log(Q)+\sqrt{\phi}\,z_{p} \right]^2 \right), 
\quad t>0,
\end{equation}
and
\begin{equation}\label{eq:quant:cd}
 F_{T}(t;Q,\phi)
=
G\!\left(\frac{1}{\phi} \left[ \log(t)-\log(\lambda) \right]^2 \right)=
G\!\left(\frac{1}{\phi} \left[ \log(t)-\log(Q)+\sqrt{\phi}\,z_{p} \right]^2 \right),\quad t>0,
\end{equation}
implying the notation $T\sim\textrm{QLS}(Q,\phi,g)$. If $T\sim\textrm{QLS}(Q,\phi,g)$, \cite{ssls:20} have shown that the following properties hold: (a) $cT\sim\textrm{QLS}(cQ,\phi,g)$, with $c>0$; (b) $T^c\sim\textrm{QLS}(Q^c,c^2\phi,g)$, with $c>0$. We then readily have the following relation:
\begin{equation}\label{eq:rela01}
T=Q\,\epsilon^{\sqrt{\phi}},  \quad \text{with}\,\,T\sim\textrm{QLS}(Q,\phi,g)\,\,\text{and}\,\,\epsilon \sim \textrm{QLS}(1, 1, g).
\end{equation}

\section{Log-symmetric quantile tobit model}\label{sec:3}
Let $T_{i}$ be a positive censored variable to the left at point $\Psi$, namely, it is observable for values greater than $\Psi$ and censored for values less than or equal to $\Psi$. Based on \eqref{eq:rela01}, the 
log-symmetric quantile tobit model can be formulated as 
\begin{eqnarray}\label{eq:tobit-logsymquantsec}
T_{i}=
\begin{cases}
\Psi, & \quad T_{i}^{\ast}  \leq  \Psi, \quad i=1,\ldots,m,\\
T_{i}^{\ast}=Q_i\epsilon_{i}^{\sqrt{\phi_i}}, & \quad  T_{i}^{\ast}  >  \Psi, \quad i=m+1,\ldots,n,
\end{cases}
\end{eqnarray}
where $\epsilon_i \sim \textrm{QLS}(1, 1, g)$, $Q_i=\exp(\bm{x}_i^\top\bm \beta)$ and 
$\phi_i =\exp(\bm{w}^{\top}_{i}\bm{\kappa})$, with $\bm{\beta} =(\beta_0,\ldots,\beta_{k})^\top$ and 
$\bm{\kappa}=(\kappa_0,\ldots,{\kappa_{l}})^\top$ denoting vectors of regression coefficients, and ${\bm{x}}^{\top}_{i}= (1,x_{i1},\ldots, x_{ik})^\top$ and 
 ${\bm{w}}^{\top}_{i} = (1,w_{i1}, \ldots, w_{il})^\top$ denoting vectors of explanatory variables fixed and known
 associated with  $Q_i$ and $\phi_i$, respectively.
 
The estimation of the parameters of the quantile log-symmetric tobit model presented in \eqref{eq:tobit-logsymquantsec} can be done by the maximum likelihood method. Let ${\bm T}=(T_1,\ldots,T_m,T_{m+1},\ldots,T_n)^{\top}$ be a sample of size $n$ from the quantile log-symmetric tobit model that contains $m$ left-censored data at $\Psi$ and $n-m$ uncensored data. Then, the corresponding log-likelihood function for the parameter vector ${\bm\theta} = ({\bm \beta}^{\top},{\bm \kappa}^{\top})^{\top}$ is given by
\begin{equation}\label{eq:like}
L({\bm\theta})= \prod_{i=1}^{m}  G\left( \frac{\log(\Psi)-\log(Q_i)+\sqrt{\phi_i}z_{p}}{\sqrt{\phi_i}} \right) \prod_{i=m+1}^{n} \frac{1}{\sqrt{\phi_i}}
g\left( \frac{[\log(t_i)-\log(Q_i)+\sqrt{\phi_i}z_{p}]^2}{\phi_i} \right),
\end{equation}
where $Q_{i}=\exp(\bm{x}_i^\top\bm \beta)$, $\phi_i=\exp(\bm{w}^{\top}_{i}\bm{\kappa})$, $G$ is as defined in \eqref{eq:lg-cdf}, and $g$ is given in Table~\ref{tab:gfuncions}. By taking the logarithm of \eqref{eq:like}, we obtain the log-likelihood function $\ell({\bm\theta})$, that is,
\begin{equation}\label{eq:loglike}
\ell({\bm\theta}) = \sum_{i=1}^{n} \ell_i({\bm\theta}),
\end{equation}
where
$$
\ell_i({\bm\theta}) =
\begin{cases}
\log\left( G\left( \frac{\log(\Psi)-\log(Q_i)+\sqrt{\phi_i}z_{p}}{\sqrt{\phi_i}} \right) \right), & i=1,\ldots,m,\\[0.25cm]
-\frac{1}{2}\log(\phi_i)+\log\left( g\left( \frac{[\log(t_i)-\log(Q_i)+\sqrt{\phi_i}z_{p}]^2}{\phi_i} \right)\right), &  i=m+1,\ldots,n.
\end{cases}
$$
By taking the first derivative of $\ell({\bm\theta})$ with respect to ${\bm \beta}$ and ${\bm \kappa}$, we obtain 
the score vector, that is,
\begin{equation}\label{eq:score}
\dot{\bm \ell}({\bm\theta})= \dfrac{\partial{\ell({\bm\theta})}}{\partial{{\bm\theta}}} = \sum_{i=1}^{n} \dot{\bm \ell}_i({\bm\theta}), \,\,
\end{equation}
where $ \dot{\bm \ell}_i({\bm\theta}) = (\dot{\bm \ell}_{i{\bm \beta}}^\top({\bm\theta}),\dot{\bm\ell}_{i{\bm \kappa}}({\bm\theta}))^\top$, with
$$
\dot{\bm\ell}_{i{\bm\beta}}({\bm\theta}) =
\begin{cases}
-\frac{1}{\sqrt{\phi_i}} \Pi(\xi^c_i) {\bm x}_{i},  & i=1,\ldots,m,\\[0.25cm]
-\frac{2}{\sqrt{\phi_i}}\Delta(\xi_i^2)\xi_{i} {\bm x}_{i} , & i=m+1,\ldots,n,
\end{cases}
$$
$$
\dot{\bm\ell}_{i{\bm\kappa}}({\bm\theta}) =
\begin{cases}
-\frac{1}{2\sqrt{\phi_i}}\Pi(\xi^c_i){\bm w_{i}}\gamma_i^c, & i=1,\ldots,m,\\[0.25cm]
-\frac{1}{2}{\bm w_{i}}-\frac{1}{\sqrt{\phi}}\Delta(\xi_i^2)\xi_{i}{\bm w_{i}} \gamma_i, & i=m+1,\ldots,n,
\end{cases}
$$
where
$\Pi(\xi^c_i)=({{\rm d}G(u)/{\rm d}u|_{u=\xi^c_i}})/{G(\xi^c_i)} $ and 
$\Delta(\xi_i^2)=({{\rm d}g(u)/{\rm d}u|_{u=\xi_{i}^2}})/{g(\xi_{i}^{2})} $, with \\
$\xi^c_i=({\log(\Psi)-\log(Q_i)+\sqrt{\phi_i}z_{p}})/{\sqrt{\phi_i}}$, 
$\xi_i=({\log(t_i)-\log(Q_i)+\sqrt{\phi_i}z_{p}})/{\sqrt{\phi_i}}$,
$\gamma_i^c=\log(\Psi)-\log(Q_i)$ and $\gamma_i=\log(t_i)-\log(Q_i)$.

The maximum likelihood estimate for $\bm{\theta}$ is obtained by maximizing the log-likelihood function \eqref{eq:loglike} by equating the score vector $\dot{\bm\ell}(\bm{\theta})$, which contains the vector of first derivatives of $\ell({\bm\theta})$, to zero, providing the likelihood equations. In this case, as there is no analytical solution, they are solved by using the Broyden-Fletcher-Goldfarb-Shanno (BFGS) iterative method for non-linear optimization. Note that in \eqref{eq:like}, the extra parameter $\xi$ is assumed to be fixed. The reason for this lies in the works of \cite{l:97} and \cite{kanoetal93}. In the first work it is shown that the robustness of the Student-$t$ distribution to outlying observations holds only when the degree of freedom parameter is fixed, instead of being estimated directly in the log-likelihood function.  In the second work, the authors report difficulties in estimating the extra parameter for the power-exponential distribution. Therefore, the parameter $\xi$ is estimated using the profiled log-likelihood. Two basic steps are required:
\begin{itemize}
 \item[S1)] Consider a grid of values $\xi_1,\xi_2,\ldots,\xi_K$. For each fixed value of $\xi_j$, $j=1,2,\ldots,K$, compute the estimate of ${\bm\theta} = ({\bm \beta}^{\top},{\bm \kappa}^{\top})^{\top}$ based on $\xi_j$, that is, $\widehat{\bm\theta}_{j} = (\widehat{\bm \beta}_{j}^{\top},\widehat{\bm \kappa}_{j}^{\top})^{\top}$. Compute also the value of the associated log-likelihood function, $\ell_{j}(\widehat{\bm\theta})$. 
\item[S2)] Obtain the final estimates of $\xi$ and ${\bm\theta} = ({\bm \beta}^{\top},{\bm \kappa}^{\top})^{\top}$, $\widehat{\xi}$ and $\widehat{\bm\theta} = (\widehat{\bm \beta}^{\top},\widehat{\bm \kappa}^{\top})^{\top}$ say, as the associated estimates that maximize the log-likelihood function ($\max_j \ell_{j}(\widehat{\bm\theta})$). 
\end{itemize}

Under regularity conditions, the asymptotic distribution of $\widehat{\bm{\theta}}$ is a multivariate normal, that is,
$$\sqrt{n}(\widehat{{\bm\theta}}-{\bm\theta})\dot{\sim}\textrm{N}_{k+l+2}\left(\bm{0}_{k+l+2}, {\bm{\Sigma}}_{{\bm{\theta}}}\right),$$
where $\,\dot{\sim}\,$ denotes convergence in distribution and ${\bm{\Sigma}}_{{\bm{\theta}}}$ is the asymptotic variance-covariance matrix of $\widehat{\bm{\theta}}$ \citep{ch:74}, which is the inverse of the expected Fisher information matrix. We can approximate the expected Fisher information matrix by its observed version obtained from the Hessian matrix $\ddot{\bm \ell}({\bm\theta})$, which contains the second derivatives of $\ell({\bm\theta})$. Thus,      
${{\bm{\Sigma}}}_{{\bm{\theta}}}\approx [-\ddot{\bm \ell}({\bm\theta})]^{-1}$, where 
\begin{equation*}\label{eq:hessian}
\ddot{\bm \ell}({\bm\theta})= \dfrac{\partial^{2}{\ell({\bm\theta})}}{\partial{{\bm\theta}}\partial{{\bm\theta}^{\top}}} = \sum_{i=1}^{n} \ddot{\bm \ell}_i({\bm\theta}), \,\,
\text{with}\,\, 
\ddot{\bm \ell}_i({\bm\theta}) =
\left[\begin{array}{cc}
\ddot{\bm\ell}_{i{\bm\beta}{\bm\beta}}({\bm\theta}) &   \ddot{\bm\ell}_{i{\bm\beta}{\bm\kappa}}({\bm\theta})\\
\ddot{\bm\ell}_{i{\bm\kappa}{\bm\beta}}({\bm\theta}) &   \ddot{\bm\ell}_{i{\bm\kappa}{\bm\kappa}}({\bm\theta})
\end{array}\right].
\end{equation*}
The elements of $\ddot{\bm \ell}({\bm\theta})$ are given by
$$
\ddot{\bm\ell}_{i{\bm\beta}{\bm\beta}}({\bm\theta}) =
\begin{cases}
\frac{1}{\phi_i} \Pi'(\xi^c_i) {\bm x}_{i}{\bm x}_{i}^{\top} ,  & i=1,\ldots,m,\\[0.25cm]
\frac{4}{\phi_i}\Delta'(\xi^2_i)\xi^2_i{\bm x_{i}}{\bm x}_{i}^{\top}+\frac{2}{\phi_i}\Delta(\xi^2_i){\bm x_{i}}{\bm x}_{i}^{\top} , & i=m+1,\ldots,n,
\end{cases}
$$
$$
\ddot{\bm\ell}_{i{\bm\beta}{\bm\kappa}}({\bm\theta}) = \ddot{\bm\ell}_{i{\bm\kappa}{\bm\beta}}({\bm\theta})=
\begin{cases}
\left[ \frac{{\bm w}_{i}}{2\sqrt{\phi_i}}\Pi(\xi^c_i)+\frac{1}{2\phi_i}\Pi'(\xi^c_i){\bm w}_{i}\gamma_i^c \right]{\bm x}_{i},  & i=1,\ldots,m,\\[0.25cm]
-2\left\{ -\frac{{\bm w}_{i}}{2\sqrt{\phi_i}}\Delta(\xi_i^2)\xi_i
-\frac{{\bm w}_{i}\gamma_i}{\phi_i}
\left[\Delta'(\xi_i^2)\xi_i^2+\frac{1}{2} \Delta(\xi_i^2) \right] \right\}{\bm x}_{i}, & i=m+1,\ldots,n,
\end{cases}
$$
$$
\dot{\bm\ell}_{i{{\bm\kappa}}{{\bm\kappa}}}({\bm\theta}) =
\begin{cases}
-\frac{1}{4\phi_i}\left[ -\Pi(\xi_i^c)+\frac{1}{\sqrt{\phi_i}}\Pi'(\xi_i^c)\gamma_i^c \right]\gamma_i^c  {\bm w}_{i}{\bm w}_{i}^{\top} , & i=1,\ldots,m,\\[0.25cm]
\left\{\frac{{\bm w}_{i}}{2\sqrt{\phi_i}}\Delta(\xi_i^2)\xi_i
+\frac{{\bm w}_{i}\gamma_i}{\phi_i}
\left[\Delta'(\xi_i^2)\xi_i^2+\frac{1}{2} \Delta(\xi_i^2) \right] \right\}{\bm w}_{i}\gamma_i, & i=m+1,\ldots,n.
\end{cases}
$$

The corresponding standard errors can then be approximated by the square roots of the diagonal elements in the variance-covariance matrix evaluated at $\widehat{\bm \theta}$.

\subsection{Interpretation of the regression coefficients}\label{sec:interp}

The regression coefficient of the proposed tobit model is interpreted in terms of the effect on the latent variable $T_i^*$ in the uncensored part. Let $\beta_j$ be the $j$-th regression coefficient and use the subscript $(j)$ to imply excluding the  $j$-th element, such that $\bm{x}_{i(j)}$ and ${\bm\beta}_{(j)}$ are, respectively, the vector of explanatory variables excluding $x_{ij}$ and the regression coefficients excluding $\beta_j$. Note that the quantile of $T_i^*$ is given by
\begin{equation*}
 Q(T_i^*|{x}_{ij},{\bm x}_{i(j)})=\exp(\beta_0+\beta_j{x}_{ij}+\bm{x}_{i(j)}^{\top}{\bm\beta}_{(j)}).
\end{equation*}
If $x_{ij}$ increases by 1 while keeping ${\bm x}_{i(j)}$ fixed, we obtain
\begin{eqnarray*}
 Q(T_i^*|{x}_{ij}+1,{\bm x}_{i(j)})&=&\exp(\beta_j(x_{ij}+1))\exp(\beta_0+\bm{x}_{i(j)}^{\top}{\bm\beta}_{(j)})\\
                                   &=&\exp(\beta_j)\exp(\beta_0+\beta_j{x}_{ij}+\bm{x}_{i(j)}^{\top}{\bm\beta}_{(j)})\\
                                   &=&\exp(\beta_j)Q(T_i^*|{x}_{ij},{\bm x}_{i(j)}).
\end{eqnarray*}
Thus, for any $j$ increasing $x_{ij}$ by 1, the quantile of $T_i^*$ will be multiplied by $\exp(\beta_j)$. This is usually expressed as a percentage change, and
\begin{equation*}
 \frac{Q(T_i^*|{x}_{ij}+1,{\bm x}_{i(j)})-Q(T_i^*|{x}_{ij},{\bm x}_{i(j)})}{Q(T_i^*|{x}_{ij},{\bm x}_{i(j)})}\times 100\%=
 (\exp(\beta_j)-1)\times 100\%
\end{equation*}
is the approximate percentage increase (or decrease if the value of $\beta_j$ is negative) in the quantile of 
$T_i^*$ when ${x}_{ij}$ is increased by 1. For $x_{ij}$ dichotomous, $(\exp(\beta_j)-1)\times 100\%$ is the percentage increase (or decrease if $\beta_j$ is negative) in the quantile of $T_i^*$ when ${x}_{ij}$ changes from 0 to 1. Note that when $-0.4\leq \beta_j \leq 0.4$, we can use the approximation $(\exp(\beta_j)-1)\approx \beta_j$; see \cite{weisberg:14}.

\subsection{Residual analysis}
Goodness of fit and departures from the assumptions of the model can be assessed through residual analysis. In this work, we work with martingale-type (MT) residual, which is given by
$$
{r}_{_{MT_{i}}}=\textrm{sign}(r_{_{\textrm{\tiny
M}_{i}}})\sqrt{-2\left( r_{_{\textrm{\tiny
M}_{i}}}+\rho_i\log(\rho_i- r_{_{\textrm{\tiny M}_{i}}})\right)},\quad i=1,\dots,n.
$$
where $r_{_{\textrm{\tiny M}_{i}}} = \rho_i + \log(\widehat S(t_i))$, with 
$\widehat S(t_i)$ being the survival function fitted to the data, and $\rho_i=0$ or $1$ indicating that case $i$ is censored or not, respectively; details on the MT residual can be seen in \citet{tgf:90}. Simulations results indicate that the empirical distribution of the MT residual is in agreement with the standard normal distribution; see \cite{SILVA20094482}. Then, a normal quantile-quantile (QQ) plot with simulated envelope can be constructed for the MT residual to verify whether the model is correctly specified.

\section{Monte Carlo simulation}\label{sec:4}

In this section, the performance of the maximum likelihood estimates of the log-symmetric quantile tobit models are evaluated by means of a Monte Carlo simulation study. The following distributions are considered: log-normal (log-NO), log-Student-$t$ (log-$t$), log-power-exponential (log-PE) and extended Birnbaum-Saunders (EBS). For each Monte Carlo replica, a simulated sample of the log-symmetric quantile tobit model is generated for fixed parameter values, then the maximum likelihood estimates are obtained for each simulated sample. Estimates of bias and mean squared error (MSE) are thus computed from the Monte Carlo replicas as
$$\widehat{\textrm{Bias}}(\widehat{\theta}) = \frac{1}{\text{NREP}} \sum_{i = 1}^{\text{NREP}} \widehat{\theta}^{(i)} - \theta
\quad \text{and}\quad
\widehat{\mathrm{MSE}}(\widehat{\theta}) = \frac{1}{\text{NREP}} \sum_{i = 1}^{\text{NREP}} (\widehat{\theta}^{(i)} - \theta)^2,$$
where $\theta$ and $\widehat{\theta}^{(i)}$ are the true parameter value and its respective $i$-th maximum likelihood estimate, and $\text{NREP}$ is the number of Monte Carlo replicas. The \texttt{R} software has been used to do all numerical
calculations; see \cite{r2020vienna}.  

Simulated data from the log-symmetric quantile tobit models are generated according to  
\begin{eqnarray}\label{eq:sim}
T_{i}=
\begin{cases}
\Psi, & \quad T_{i}^{\ast}  \leq  \Psi, \quad i=1,\ldots,m,\\
T_{i}^{\ast}=Q_i\epsilon_{i}^{\sqrt{\phi_i}}, & \quad  T_{i}^{\ast}  >  \Psi, \quad i=m+1,\ldots,n,
\end{cases}
\end{eqnarray}
where $\epsilon_i \sim \textrm{QLS}(1, 1, g)$, $Q_i=\exp(\beta_0 + \beta_1 x_i)$ and $\phi_i=\exp(\kappa_0 + \kappa_1 w_i)$.  

The simulation scenario considers: $\beta_0 = 1.0$, $\beta_1 = 0.5$, $\kappa_0 = 1$, $\kappa_1 = 1.5$,  $q = 0.10, 0.50, 0.90$ and $\text{NREP}=5,000$. The explanatory variables $x_i$ and $w_i$ are generated from the Bernoulli(0.5) and Uniform(0,1) distributions, respectively. The values of the extra parameters for the distributions considered are: $\xi=4$ (log-$t$), $\xi=0.3$ (log-PE) and $\xi=0.5$ (EBS). The value of $\Psi$ in \eqref{eq:sim} is determined so that the censoring proportion is 10\% or 40\%.

Tables \ref{tab:sim:01}-\ref{tab:sim:04} present the results of Monte Carlo simulations based on the log-NO, log-$t$, log-PE and EBS distributions. These tables report the bias and MSE obtained for different combinations of censoring proportion, $q$ and sample size ($n$). A look at the results in Tables \ref{tab:sim:01}-\ref{tab:sim:04} allows us to conclude that, in general, as the sample size increases, the bias and MSE both decrease, as expected. This results is expected since the maximum likelihood estimator is consistent \citep{Greene2003Econometric}. Moreover, we observe that, in general, when the censoring proportion increases, the bias and MSE both increase, that is, the performances of the estimates deteriorates, a result also expected since the likelihood function loses information contained in the sample when the percentage of censoring increases \citep{W-1992}.

\begin{table}[!ht]
	\scriptsize
	\centering
	\caption{Bias and MSE from simulated data for the log-NO quantile tobit model.}
		\adjustbox{max height=\dimexpr\textheight-3.5cm\relax,
		max width=\textwidth}{
	\begin{tabular}{cccccccccccccc}
	\toprule
\multirow{1}{*}{Censoring}&\multirow{1}{*}{$q$}& Parameter& \multicolumn{2}{c}{$n = 50$} & &\multicolumn{2}{c}{$n = 300$} & &\multicolumn{2}{c}{$n = 600$}\\
 	\cline{4-5} \cline{7-8} \cline{10-11}
                    &	&                                & Bias   & MSE     & & Bias   & MSE     & & Bias   & MSE\\
	\hline
\multirow{12}{*}{10\%}&\multirow{4}{*}{0.10}& $\beta_0$  & 0.0189 & 0.3423  & &-0.0042 & 0.0476  & &-0.0071 & 0.0288 \\
                      &                     & $\beta_1$  & 0.0162 & 0.4733  & & 0.0080 & 0.0693  & & 0.0090 & 0.0366 \\
                      &                     & $\kappa_0$ &-0.0613 & 0.1729  & &-0.0011 & 0.0205  & &-0.0031 & 0.0119 \\
                      &                     & $\kappa_1$ & 0.0351 & 0.3493  & &-0.0031 & 0.0502  & & 0.0018 & 0.0245 \\[0.3ex]
                      &\multirow{4}{*}{0.50}& $\beta_0$  &-0.0151 & 0.2224  & &-0.0033 & 0.0289  & &-0.0083 & 0.0189   \\
                      &                     & $\beta_1$  & 0.0149 & 0.4646  & & 0.0067 & 0.0690  & & 0.0088 & 0.0365   \\
                      &                     & $\kappa_0$ &-0.1039 & 0.2545  & &-0.0043 & 0.0268  & &-0.0046 & 0.0155   \\
                      &                     & $\kappa_1$ & 0.1336 & 0.7881  & & 0.0062 & 0.0984  & & 0.0064 & 0.0463   \\[0.3ex]
											&\multirow{4}{*}{0.90}& $\beta_0$  &-0.0550 & 0.3099  & &-0.0025 & 0.0412  & &-0.0090 & 0.0268 \\
                      &                     & $\beta_1$  &-0.0017 & 0.4572  & & 0.0061 & 0.0692  & & 0.0086 & 0.0365 \\
                      &                     & $\kappa_0$ &-0.1221 & 0.1790  & &-0.0097 & 0.0199  & &-0.0057 & 0.0114 \\
                      &                     & $\kappa_1$ & 0.1623 & 0.4417  & & 0.0160 & 0.0561  & & 0.0078 & 0.0275 \\[0.8ex]     \cline{1-11}                
\multirow{12}{*}{40\%}&\multirow{4}{*}{0.10}& $\beta_0$  & 0.0596 & 0.5899  & & 0.0080 & 0.0866  & &-0.0062 & 0.0503 \\
                      &                     & $\beta_1$  & 0.0133 & 0.5460  & & 0.0113 & 0.0769  & & 0.0112 & 0.0436 \\
                      &                     & $\kappa_0$ &-0.1130 & 0.2639  & &-0.0131 & 0.0315  & &-0.0068 & 0.0172 \\
                      &                     & $\kappa_1$ & 0.0637 & 0.3987  & & 0.0043 & 0.0552  & & 0.0049 & 0.0268 \\[0.3ex]
											&\multirow{4}{*}{0.50}& $\beta_0$  &-0.0205 & 0.2643  & &-0.0047 & 0.0349  & &-0.0097 & 0.0233 \\
                      &                     & $\beta_1$  & 0.0018 & 0.5243  & & 0.0105 & 0.0759  & & 0.0075 & 0.0415 \\
                      &                     & $\kappa_0$ &-0.0954 & 0.3909  & &-0.0057 & 0.0456  & &-0.0018 & 0.0243 \\
                      &                     & $\kappa_1$ & 0.0619 & 1.1719  & &-0.0051 & 0.1562  & &-0.0011 & 0.0692 \\[0.3ex]
											&\multirow{4}{*}{0.90}& $\beta_0$  &-0.0654 & 0.3359  & &-0.0066 & 0.0444  & &-0.0077 & 0.0286 \\
                      &                     & $\beta_1$  & 0.0054 & 0.5160  & & 0.0115 & 0.0761  & & 0.0066 & 0.0410 \\
                      &                     & $\kappa_0$ &-0.1076 & 0.2590  & &-0.0029 & 0.0317  & &-0.0012 & 0.0163 \\
                      &                     & $\kappa_1$ & 0.1552 & 0.7737  & &-0.0027 & 0.1086  & & 0.0014 & 0.0481 \\ 
                     
	\bottomrule
	\end{tabular}}
\label{tab:sim:01}
\end{table}

\begin{table}[!ht]
	\scriptsize
	\centering
	\caption{Bias and MSE from simulated data for the log-$t$ quantile tobit model.}
		\adjustbox{max height=\dimexpr\textheight-3.5cm\relax,
		max width=\textwidth}{
	\begin{tabular}{cccccccccccccc}
	\toprule
\multirow{1}{*}{Censoring}&\multirow{1}{*}{$q$}& Parameter& \multicolumn{2}{c}{$n = 50$} & &\multicolumn{2}{c}{$n = 300$} & &\multicolumn{2}{c}{$n = 600$}\\
 	\cline{4-5} \cline{7-8} \cline{10-11}
                    &	&                                & Bias   & MSE     & & Bias   & MSE     & & Bias   & MSE\\
	\hline
\multirow{12}{*}{10\%}&\multirow{4}{*}{0.10}& $\beta_0$  & 0.1100 & 0.5640  & & 0.0149 & 0.0855  & & 0.0021 & 0.0441 \\
                      &                     & $\beta_1$  & 0.0266 & 0.6622  & & 0.0059 & 0.1019  & &-0.0017 & 0.0458 \\
                      &                     & $\kappa_0$ &-0.1338 & 0.2383  & &-0.0147 & 0.0341  & &-0.0077 & 0.0185 \\
                      &                     & $\kappa_1$ & 0.1138 & 0.4357  & & 0.0016 & 0.0692  & & 0.0094 & 0.0345 \\[0.3ex]
											
                      &\multirow{4}{*}{0.50}& $\beta_0$  & 0.0067 & 0.3156  & &-0.0027 & 0.0449  & &-0.0012 & 0.0257 \\
                      &                     & $\beta_1$  & 0.0136 & 0.6573  & & 0.0044 & 0.1011  & &-0.0021 & 0.0457 \\
                      &                     & $\kappa_0$ &-0.1777 & 0.4093  & &-0.0175 & 0.0484  & &-0.0038 & 0.0277 \\
                      &                     & $\kappa_1$ & 0.2236 & 1.2232  & & 0.0105 & 0.1581  & & 0.0017 & 0.0819 \\[0.3ex]
											
                      &\multirow{4}{*}{0.90}& $\beta_0$  &-0.1133 & 0.5224  & &-0.0212 & 0.0819  & &-0.0051 & 0.0469 \\
                      &                     & $\beta_1$  & 0.0038 & 0.6644  & & 0.0048 & 0.1030  & &-0.0023 & 0.0456 \\
                      &                     & $\kappa_0$ &-0.1656 & 0.2597  & &-0.0230 & 0.0327  & &-0.0027 & 0.0169 \\
                      &                     & $\kappa_1$ & 0.1738 & 0.4910  & & 0.0217 & 0.0677  & &-0.0023 & 0.0312 \\[0.8ex] \cline{1-11}                     
\multirow{12}{*}{40\%}&\multirow{4}{*}{0.10}& $\beta_0$  & 0.0829 & 0.9005  & & 0.0148 & 0.1392  & & 0.0033 & 0.0771 \\
                      &                     & $\beta_1$  & 0.0475 & 0.7609  & & 0.0089 & 0.1115  & & 0.0032 & 0.0516 \\
                      &                     & $\kappa_0$ &-0.1744 & 0.3822  & &-0.0225 & 0.0512  & &-0.0141 & 0.0273 \\
                      &                     & $\kappa_1$ & 0.1506 & 0.5333  & & 0.0073 & 0.0799  & & 0.0144 & 0.0386 \\[0.3ex]
											
                      &\multirow{4}{*}{0.50}& $\beta_0$  &-0.0198 & 0.3453  & &-0.0070 & 0.0488  & &-0.0052 & 0.0281 \\
                      &                     & $\beta_1$  & 0.0245 & 0.7086  & & 0.0056 & 0.1068  & & 0.0001 & 0.0488 \\
                      &                     & $\kappa_0$ &-0.1651 & 0.6005  & &-0.0129 & 0.0785  & &-0.0048 & 0.0404 \\
                      &                     & $\kappa_1$ & 0.1960 & 1.8149  & &-0.0045 & 0.2635  & & 0.0048 & 0.1207 \\[0.3ex]
											
                      &\multirow{4}{*}{0.90}& $\beta_0$  &-0.1155 & 0.5832  & &-0.0226 & 0.0918  & &-0.0062 & 0.0513 \\
                      &                     & $\beta_1$  & 0.0252 & 0.7256  & & 0.0085 & 0.1098  & &-0.0024 & 0.0480 \\
                      &                     & $\kappa_0$ &-0.1560 & 0.3205  & &-0.0236 & 0.0411  & &-0.0050 & 0.0211 \\
                      &                     & $\kappa_1$ & 0.1876 & 0.7113  & & 0.0308 & 0.1029  & & 0.0036 & 0.0500 \\ 
                     
	\bottomrule
	\end{tabular}}
\label{tab:sim:02}
\end{table}

\begin{table}[!ht]
	\scriptsize
	\centering
	\caption{Bias and MSE from simulated data for the log-PE quantile tobit model.}
		\adjustbox{max height=\dimexpr\textheight-3.5cm\relax,
		max width=\textwidth}{
	\begin{tabular}{cccccccccccccc}
	\toprule
\multirow{1}{*}{Censoring}&\multirow{1}{*}{$q$}& Parameter& \multicolumn{2}{c}{$n = 50$} & &\multicolumn{2}{c}{$n = 300$} & &\multicolumn{2}{c}{$n = 600$}\\
 	\cline{4-5} \cline{7-8} \cline{10-11}
                    &	&                                & Bias   & MSE     & & Bias   & MSE     & & Bias   & MSE\\
	\hline
\multirow{12}{*}{10\%}&\multirow{4}{*}{0.10}& $\beta_0$  & 0.0584 & 0.6195  & & 0.0184 & 0.0860  & &-0.0021 & 0.0465 \\
                      &                     & $\beta_1$  &-0.0071 & 0.7409  & &-0.0046 & 0.1065  & &-0.0010 & 0.0556 \\
                      &                     & $\kappa_0$ &-0.0816 & 0.1918  & &-0.0134 & 0.0265  & &-0.0016 & 0.0148 \\
                      &                     & $\kappa_1$ & 0.0525 & 0.3697  & & 0.0107 & 0.0626  & &-0.0017 & 0.0277 \\[0.3ex]
											
                      &\multirow{4}{*}{0.50}& $\beta_0$  &-0.0114 & 0.3690  & & 0.0067 & 0.0503  & &-0.0045 & 0.0283 \\
                      &                     & $\beta_1$  &-0.0029 & 0.7270  & &-0.0058 & 0.1059  & &-0.0005 & 0.0559 \\
                      &                     & $\kappa_0$ &-0.1004 & 0.3137  & &-0.0206 & 0.0380  & &-0.0060 & 0.0214 \\
                      &                     & $\kappa_1$ & 0.1039 & 0.9654  & & 0.0314 & 0.1413  & & 0.0104 & 0.0636 \\[0.3ex]
											
                      &\multirow{4}{*}{0.90}& $\beta_0$  &-0.0851 & 0.5457  & &-0.0029 & 0.0770  & &-0.0050 & 0.0459 \\
                      &                     & $\beta_1$  &-0.0145 & 0.7241  & &-0.0053 & 0.1063  & &-0.0008 & 0.0560 \\
                      &                     & $\kappa_0$ &-0.1117 & 0.2157  & &-0.0198 & 0.0247  & &-0.0089 & 0.0139 \\
                      &                     & $\kappa_1$ & 0.1237 & 0.4721  & & 0.0292 & 0.0656  & & 0.0161 & 0.0301 \\[0.8ex] \cline{1-11}                     
\multirow{12}{*}{40\%}&\multirow{4}{*}{0.10}& $\beta_0$  & 0.0864 & 0.9524  & & 0.0319 & 0.1411  & &-0.0022 & 0.0767 \\
                      &                     & $\beta_1$  & 0.0006 & 0.8329  & &-0.0056 & 0.1160  & &-0.0013 & 0.0629 \\
                      &                     & $\kappa_0$ &-0.1354 & 0.3078  & &-0.0247 & 0.0416  & &-0.0051 & 0.0226 \\
                      &                     & $\kappa_1$ & 0.0923 & 0.4657  & & 0.0154 & 0.0717  & & 0.0016 & 0.0319 \\[0.3ex]
											
                      &\multirow{4}{*}{0.50}& $\beta_0$  &-0.0263 & 0.4142  & & 0.0067 & 0.0544  & &-0.0046 & 0.0317 \\
                      &                     & $\beta_1$  & 0.0075 & 0.8002  & &-0.0072 & 0.1109  & &-0.0017 & 0.0596 \\
                      &                     & $\kappa_0$ &-0.1100 & 0.4897  & &-0.0214 & 0.0575  & &-0.0123 & 0.0330 \\
                      &                     & $\kappa_1$ & 0.0598 & 1.5340  & & 0.0143 & 0.2073  & & 0.0179 & 0.0937 \\[0.3ex]
											
                      &\multirow{4}{*}{0.90}& $\beta_0$  &-0.1125 & 0.6255  & &-0.0078 & 0.0845  & &-0.0040 & 0.0494 \\
                      &                     & $\beta_1$  & 0.0035 & 0.7936  & &-0.0062 & 0.1156  & &-0.0010 & 0.0593 \\
                      &                     & $\kappa_0$ &-0.1164 & 0.3024  & &-0.0208 & 0.0335  & &-0.0112 & 0.0186 \\
                      &                     & $\kappa_1$ & 0.1342 & 0.7742  & & 0.0255 & 0.0987  & & 0.0260 & 0.0464 \\ 
                     
	\bottomrule
	\end{tabular}}
\label{tab:sim:03}
\end{table}

\begin{table}[t]
\scriptsize
	\centering
	\caption{Bias and MSE from simulated data for the EBS quantile tobit model.}
		\adjustbox{max height=\dimexpr\textheight-3.5cm\relax,
		max width=\textwidth}{
	\begin{tabular}{cccccccccccccc}
	\toprule
\multirow{1}{*}{Censoring}&\multirow{1}{*}{$q$}& Parameter& \multicolumn{2}{c}{$n = 50$} & &\multicolumn{2}{c}{$n = 300$} & &\multicolumn{2}{c}{$n = 600$}\\
 	\cline{4-5} \cline{7-8} \cline{10-11}
                    &	&                                & Bias   & MSE     & & Bias   & MSE     & & Bias   & MSE\\
	\hline
\multirow{12}{*}{10\%}&\multirow{4}{*}{0.10}& $\beta_0$  & 0.0029 & 0.0204  & &-0.0016 & 0.0029  & &-0.0021 & 0.0018 \\
                      &                     & $\beta_1$  & 0.0111 & 0.0278  & & 0.0032 & 0.0041  & & 0.0027 & 0.0022 \\
                      &                     & $\kappa_0$ &-0.0742 & 0.1628  & &-0.0019 & 0.0188  & &-0.0033 & 0.0109 \\
                      &                     & $\kappa_1$ & 0.0466 & 0.3341  & &-0.0035 & 0.0466  & & 0.0019 & 0.0226 \\[0.3ex]
											
                      &\multirow{4}{*}{0.50}& $\beta_0$  &-0.0070 & 0.0134  & &-0.0020 & 0.0017  & &-0.0025 & 0.0011 \\
                      &                     & $\beta_1$  & 0.0093 & 0.0276  & & 0.0032 & 0.0041  & & 0.0029 & 0.0022 \\
                      &                     & $\kappa_0$ &-0.1176 & 0.2404  & &-0.0006 & 0.0242  & &-0.0045 & 0.0143 \\
                      &                     & $\kappa_1$ & 0.1477 & 0.7440  & &-0.0036 & 0.0898  & & 0.0051 & 0.0422 \\[0.3ex]
											
                      &\multirow{4}{*}{0.90}& $\beta_0$  &-0.0180 & 0.0179  & &-0.0014 & 0.0024  & &-0.0026 & 0.0015 \\
                      &                     & $\beta_1$  & 0.0041 & 0.0270  & & 0.0023 & 0.0041  & & 0.0028 & 0.0022 \\
                      &                     & $\kappa_0$ &-0.1319 & 0.1723  & &-0.0099 & 0.0184  & &-0.0057 & .0105 \\
                      &                     & $\kappa_1$ & 0.1742 & 0.4400  & & 0.0153 & 0.0540  & & 0.0082 & 0.0263 \\[0.8ex] \cline{1-11}                     
\multirow{12}{*}{40\%}&\multirow{4}{*}{0.10}& $\beta_0$  & 0.0088 & 0.0397  & & 0.0015 & 0.0059  & &-0.0028 & 0.0035 \\
                      &                     & $\beta_1$  & 0.0046 & 0.0348  & & 0.0021 & 0.0049  & & 0.0030 & 0.0028 \\
                      &                     & $\kappa_0$ &-0.0999 & 0.2275  & &-0.0120 & 0.0282  & &-0.0047 & 0.0156 \\
                      &                     & $\kappa_1$ & 0.0504 & 0.3668  & & 0.0035 & 0.0507  & & 0.0033 & 0.0249 \\[0.3ex]
											
                      &\multirow{4}{*}{0.50}& $\beta_0$  &-0.0059 & 0.0186  & &-0.0013 & 0.0024  & &-0.0025 & 0.0016 \\
                      &                     & $\beta_1$  & 0.0011 & 0.0333  & & 0.0023 & 0.0047  & & 0.0020 & 0.0027 \\
                      &                     & $\kappa_0$ &-0.1010 & 0.3638  & &-0.0054 & 0.0419  & &-0.0022 & 0.0231 \\
                      &                     & $\kappa_1$ & 0.0585 & 1.1172  & &-0.0075 & 0.1393  & &-0.0017 & 0.0637 \\[0.3ex]
											
                      &\multirow{4}{*}{0.90}& $\beta_0$  &-0.0215 & 0.0197  & &-0.0018 & 0.0025  & &-0.0025 & 0.0016 \\
                      &                     & $\beta_1$  & 0.0058 & 0.0318  & & 0.0026 & 0.0046  & & 0.0019 & 0.0025 \\
                      &                     & $\kappa_0$ &-0.1075 & 0.2481  & &-0.0021 & 0.0311  & &-0.0026 & 0.0162 \\
                      &                     & $\kappa_1$ & 0.1429 & 0.7743  & &-0.0033 & 0.1032  & &-0.0005 & 0.0491 \\ 
                     
	\bottomrule
	\end{tabular}}
\label{tab:sim:04}
\end{table}

\clearpage
\section{Application to real data}\label{sec:5}
In this section, the proposed log-symmetric quantile tobit models are used to analyze the PSID and PNAD data. The PSID data set has already been analyzed in the literature on tobit models, whereas the PNAD data set was filtered by the authors from the original data available at the IBGE website\footnote{https://www.ibge.gov.br/.}. The distributions considered are the same used in the Monte Carlo simulation.

\subsection{PSID data}

This application considers a data set corresponding to the PSID of 1976, based on data from the previous year, 1975; see 
\cite{m:87}. This set contains $n=753$ observations of white married women between 30 and 60 years of age in 1975 (year of the interview was in 1976). Of these 753 women, 325 have a salary equal to zero, that is, censored at zero. Since the proposed models are for positive data, the dependent variable is considered to be $ T + 1 $, such that $\Psi=1$.

The objective here is to study, using log-symmetric quantile tobit models, the labor supply of white married women. The dependent variable is the hourly wage ($T$) (in 1975 US dollars) and the explanatory variables are: $age$, age in years; 
$educ$, education in years; $chil6$, number of children under 6 in the household; $chil618$, number of children between 6 and 18 years old in the household; and $exper$, years of previous experience in the labor market. These data were previously studied by \cite{bgls:17} using the classic normal tobit model and the Student-$t$ tobit model.

\begin{figure}[!ht]
\centering
\subfigure[]{\includegraphics[height=7cm,width=7cm]{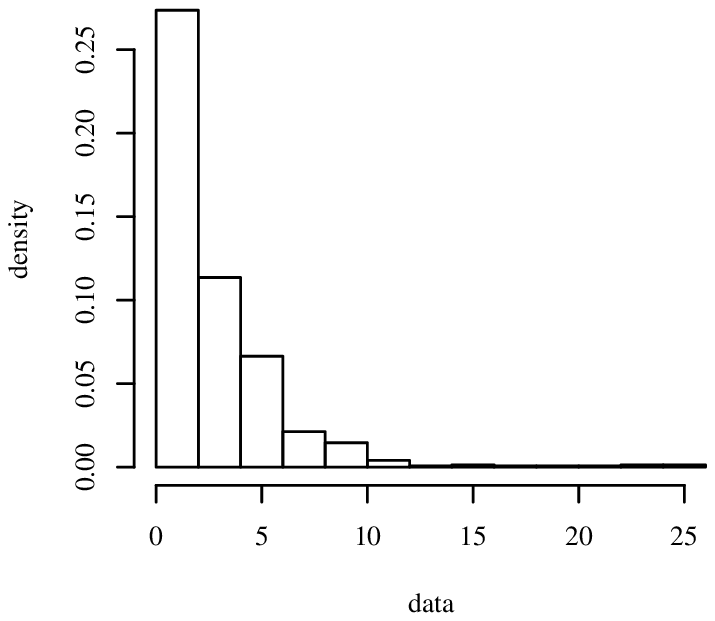}}
\subfigure[]{\includegraphics[height=6cm,width=6cm]{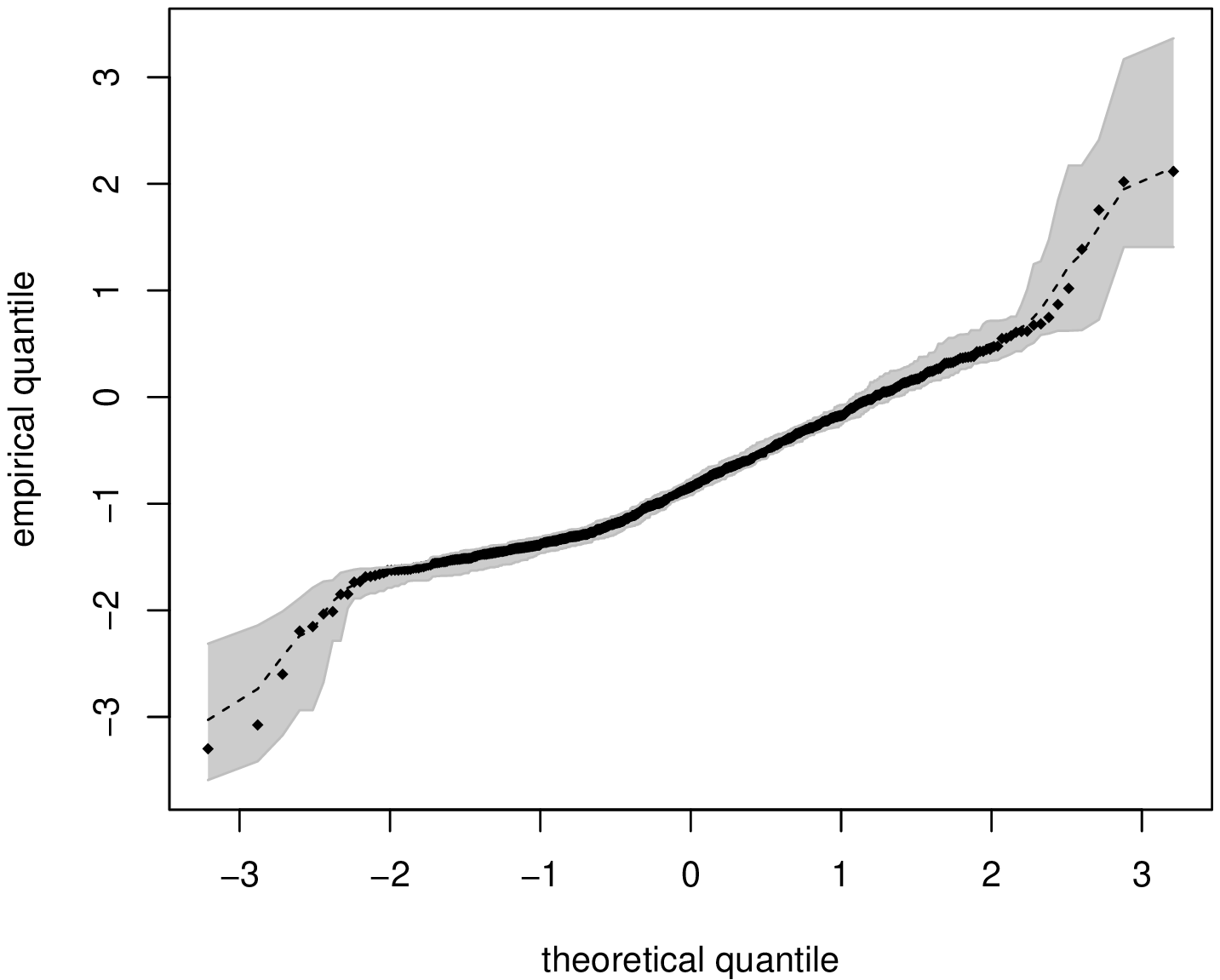}}
%\vspace{-0.25cm}
 \caption{\small {Histogram (a) for the observed women's hourly wages and QQ plot (b) and its envelope for the MT residual for the log-PE quantile tobit model ($q=0.50$).}}
\label{fig:histqqplot1}
\end{figure}

Descriptive statistics of the observed women's hourly wages reveal that the mean is equal to 2.375 and the median is equal to 1.625, that is, the mean is greater than the median. Moreover, the coefficient of variation (dispersion around the mean) is 136.52\%, indicating a high dispersion of data around the mean. Finally, the coefficients of skewness and kurtosis are equal to 2.772 and 12.755, respectively. The result of the coefficient of skewness indicates the presence of positive skewness and the coefficient of kurtosis indicates the presence of heavy tails. The asymmetric nature of the data is confirmed by the histogram shown in Figure \ref{fig:histqqplot1}(a). Such results make the proposed log-symmetric quantile tobit models good candidates, since they are models for asymmetric data, with or without heavy tails. On the other hand, the classic normal tobit model and the Student- $t$ tobit model studied by \cite{bgls:17} are not suitable when positive skewness is present.

The proposed models can accommodate heteroscedasticity, then two versions are considered:
\begin{eqnarray*}
T_{i}=
\begin{cases}
\Psi, & \quad T_{i}^{\ast}  \leq  \Psi, \quad i=1,\ldots,325,\\
T_{i}^{\ast}=Q_i\epsilon_{i}^{\sqrt{\phi_i}}, & \quad  T_{i}^{\ast}  >  \Psi, \quad i=326,\ldots,753,
\end{cases}
\end{eqnarray*}
where
$$Q_i=(\beta_0 + \beta_1\, age_i + \beta_2\, educ_i + \beta_3\, chil6_i +\beta_4\, chil618_i +  \beta_5\, exper_i),$$
$\epsilon_i \sim \textrm{QLS}(1, 1, g)$, and
\begin{itemize}
\item \textbf{Model 1}: $\phi_i=\exp\left(\kappa_0\right)$.
\item \textbf{Model 2}: $\phi_i=\exp\left( \kappa_0 + \kappa_1\, age_i + \kappa_2\, educ_i + \kappa_3\, chil6_i + \kappa_4\, chil618_i +  \kappa_5\, exper_i\right)$.
\end{itemize}
Note that the difference between the two models lies in the insertion of explanatory variables in the dispersion parameter 
$\phi$. Thus, Model 2 considers the presence of heteroscedasticity.

Table \ref{tab:resultadosaicbic} reports the AIC and BIC values for the adjusted log-symmetric quantile tobit models with 
$q=\{0.05,0.25,0.50,0.75,0.95\}$; similar results are obtained considering $ q = \{0.01, 0.02, \ldots, 0.99\}$. From this table, we observe that the models with explanatory variables in the dispersion parameter ($\phi$) provide better adjustments compared to those without explanatory variables, based on the values of AIC and BIC. In general, the results of Table \ref{tab:resultadosaicbic} indicate that the log-PE quantile tobit model presents the best fit, a result that is corroborated by the QQ plot with simulated envelope of the MT residual for this model with $q = 0.50$ (similar adjustments are found for other values of $q$), shown in Figure ~\ref{fig:histqqplot1}(b).

We can compare the proposed models to the classic normal tobit model and the Student-$t$ tobit model, which were fitted by \cite{bgls:17} using the same data set. The authors obtained $\text{AIC}=2893.85$ and $\text{BIC}=2926.22$ for the normal case, and $\text{AIC}=2760.99$ e $\text{BIC}=2793.36$ for the Student-$t$ case. Thus, we highlight the superiority of the proposed models when compared to the normal classic and Student-$t$ models adjusted by \cite{bgls:17} under three basic aspects: (i) the use of asymmetric distributions (log-symmetric distributions) that is more adequate for the PSID data. The normal classic and Student-$t$ tobit models use symmetric distributions; (ii) the possibility of modeling the dispersion and the consequent accommodation of heteroscedasticity, which improves the fit; and (iii) the modeling in terms of quantiles, which provides a richer characterization of the effects of the explanatory variables on the dependent variable. The normal classic tobit and Student-$t$ models do not consider the quantile approach.

% latex table generated in R 3.6.3 by xtable 1.8-4 package
% Sun Jun  7 18:10:05 2020
\begin{table}[!ht]
\footnotesize
\centering
 \caption{\small {AIC and BIC values for different models and $q$ with the PSID data.}}
\begin{tabular}{lrrrrrrrrrrrrrr}
  \hline
           &Model    &  Criterion  &$q=0.05$  & $q=0.25$ & $q=0.50$ & $q=0.75$ & $q=0.95$ \\ 
					\hline
Log-NO     &  1        &  AIC       & 1776.663 & 1776.663 & 1776.663 & 1776.663 & 1776.663  \\        
           &  1        &  BIC       & 1809.031 & 1809.031 & 1809.031 & 1809.031 & 1809.031  \\  
           &  2        &  AIC       & 1726.660 & 1719.173 & 1712.621 & 1705.087 & 1694.797  \\       
           &  2        &  BIC       & 1782.149 & 1774.662 & 1768.110 & 1760.576 & 1750.286  \\[0.15cm] 
Log-$t$    &  1        &  AIC       & 1776.414 & 1776.414 & 1776.414 & 1776.414 & 1776.414  \\        
           &  1        &  BIC       & 1808.783 & 1808.783 & 1808.783 & 1808.783 & 1808.783  \\  
           &  2        &  AIC       & 1727.511 & 1719.861 & 1713.169 & 1705.417 & 1688.396  \\       
           &  2        &  BIC       & 1783.000 & 1775.350 & 1768.658 & 1760.905 & 1743.885  \\ [0.15cm]
Log-PE     &  1        &  AIC       & 1775.282 & 1775.282 & 1775.282 & 1775.282 & 1775.282  \\        
           &  1        &  BIC       & 1807.651 & 1807.651 & 1807.651 & 1807.651 & 1807.651  \\  
           &  2        &  AIC       & 1698.256 & 1691.372 & 1683.956 & 1676.217 & 1669.400  \\       
           &  2        &  BIC       & 1753.745 & 1746.860 & 1739.445 & 1731.706 & 1724.889  \\[0.15cm] 
EBS        &  1        &  AIC       & 1776.691 & 1776.691 & 1776.691 & 1776.691 & 1776.691  \\        
           &  1        &  BIC       & 1809.060 & 1809.060 & 1809.060 & 1809.060 & 1809.060  \\  
           &  2        &  AIC       & 1719.489 & 1712.854 & 1706.894 & 1700.366 & 1692.420  \\       
           &  2        &  BIC       & 1774.978 & 1768.342 & 1762.383 & 1755.855 & 1747.909  \\ 										
 \hline
\end{tabular}
\label{tab:resultadosaicbic}
\end{table}

Table \ref{tab:resultadosquantis} reports the maximum likelihood estimates and standard errors for the log-PE quantile tobit model parameters based on Model 2, considering $ q =\{0.05, 0.25, 0.50, 0.75, 0.95\}$. This table also presents the estimation results based on the  optimal quantile, denoted by $q_{otm}$, which was chosen through a profile approach, that is, for a grid of values of $ q=\{0.01, 0.02, \ldots, 0.99 \} $, we estimated the model parameters and computed the corresponding AIC and BIC values. Then, the value of $q_{otm}$ was the one which had the lowest AIC and BIC values. From 
Table \ref{tab:resultadosquantis}, we note that the maximum likelihood estimates of the model parameters change according to the value of $q$, that is, the magnitude of the effect of the explanatory variables varies with $q$. We can interpret the estimated coefficients in terms of the effect on the latent variable $T_i^*$, that is, in terms of the effect on the observed part of the hourly wage; see Subsection \ref{sec:interp}. We observe, for example, that an increase of 1 year in the experience ($exper$), increases in $(\exp(0.0974)-1)*100\%=10.23\%$ the $5^{\circ}$ percentile ($q=0.05$) of the hourly wage, while increasing by $(\exp(0.0383)-1)*100\%=3.90\%$ the $75^{\circ}$ percentile ($q=0.75$) of hourly wages. In other words, the effect of increased experience on the observed part of the hourly wage is greater for women with lower income (lower quantiles). From Table \ref{tab:resultadosquantis} we also observe that the parameter estimates associated with the explanatory variables $age$, $chil6$, $exper$, which model the dispersion, are significant, indicating the presence of heteroscedasticity in the data, justifying the dispersion modeling.

% latex table generated in R 3.6.3 by xtable 1.8-4 package
% Sun Jun  7 18:10:05 2020
\begin{table}[t]
\footnotesize
\centering
 \caption{\small {Maximum likelihood estimates (standard errors in parentheses) for the log-PE quantile tobit model across different values of $q$ (Model 2).}}
\begin{tabular}{lrrrrrrrrrrrrrrrrr}
  \hline
 & $q=0.05$ & $q=0.25$ & $q=0.50$ & $q=0.75$ & $q=0.95$  & $q_{otm}=0.99$\\ 
  \hline
$\beta_0\,(intercept)$ & 0.2486   & 0.4507   & 0.5790 & 0.6487** & 0.7147**   & 0.7921 \\ 
                        & (0.6830) & (0.5235) & (0.3964) & (0.3554) & (0.4319) & (0.5196)\\ 
$\beta_1\,(age)$      & -0.1070*  & -0.0864* & -0.0631* & -0.0350* & -0.0013 & 0.0159**\\ 
                        & (0.0111) & (0.0089) & (0.0071) & (0.0064) & (0.0078) & (0.0093)\\ 
$\beta_2\,(educ)$       & 0.1708*   & 0.1638* & 0.1537* & 0.1393* & 0.1213*    & 0.1124*\\ 
                        & (0.0386) & (0.0293) & (0.0212) & (0.0167) & (0.0191) & (0.0227)\\  
$\beta_3\,(chil6)$      & -1.7487*  & -1.4588* & -1.1038* & -0.6507* & -0.0929 & 0.1946\\ 
                        & (0.2786) & (0.2301) & (0.1785) & (0.1354) & (0.1588) & (0.1985)\\
$\beta_4\,(chil618)$    & -0.0314  & -0.0113 & 0.0017 & 0.0030 & -0.0152       & -0.0320 \\ 
                        & (0.0613) & (0.0475) & (0.0362) & (0.0313) & (0.0377) &  (0.0445)\\  
$\beta_5\,(exper)$      & 0.0974*   & 0.0814* & 0.0625* & 0.0383* & 0.0061     & -0.0123 \\ 
                        & (0.0091) & (0.0074) & (0.0062) & (0.0061) & (0.0079) & (0.0093) \\[0.30cm]
$\kappa_0\,(intercept)$ & -0.9522** & -1.0583* & -1.1658* & -1.2561* & -1.2605*& -1.1975*\\ 
                         & (0.5074) & (0.5335) & (0.5619) & (0.5809) & (0.5814) & (0.5912)\\  
$\kappa_1\,(age)$      & 0.0577* & 0.0632* & 0.0695* & 0.0760* & 0.0800*      & 0.0794*\\  
                         & (0.0074) & (0.0081) & (0.0089) & (0.0097) & (0.0101) & (0.0102)\\  
$\kappa_2\,(educ)$       & -0.0249 & -0.0298 & -0.0352 & -0.0401 & -0.0415      & -0.0400\\ 
                         & (0.0262) & (0.0273) & (0.0283) & (0.0290) & (0.0284) & (0.0279)\\  
$\kappa_3\,(chil6)$      & 0.8337* & 0.9299* & 1.0337* & 1.1305* & 1.1777*      & 1.1655* \\ 
                         & (0.1434) & (0.1584) & (0.1752) & (0.1893) & (0.1944) & (0.1919)\\ 
$\kappa_4\,(chil618)$    & 0.0401 & 0.0319 & 0.0194 & -0.0003 & -0.0282         & -0.0417\\ 
                         & (0.0423) & (0.0448) & (0.0475) & (0.0504) & (0.0521) & (0.0521)\\ 
$\kappa_5\,(exper)$      & -0.0445* & -0.0497* & -0.0560* & -0.0638* & -0.0720* & -0.0750*\\
                         & (0.0061) & (0.0067) & (0.0074) & (0.0082) & (0.0090) & (0.0091)\\  
$\xi$                    & -0.48 & -0.48 & -0.48 &-0.48 &-0.48 & -0.48\\   
    \hline \multicolumn{6}{l}{\scriptsize{Source: Elaborated by the authors based on PSID data.}}\\[-0.1cm]
        \multicolumn{6}{l}{\scriptsize{* significant at 5\% level. ** significant at 10\% level.}}
\end{tabular}
\label{tab:resultadosquantis}
\end{table}

\subsection{PNAD data}

In this subsection, the log-symmetric quantile tobit models are illustrated using data from the PNAD for the year 2015, from the IBGE, which reports demographic and socioeconomic characteristics of the Brazilian population annually. A sample of the PNAD will be used, composed only of women\footnote{The most recent Continuous PNAD data will not be used, as in its dictionary there is a lack of objectivity in some variables of interest for this work, such as the experience/skill variable, which are the years of work in the main activity, as well as the variable marital status, which is also not clearly defined in the Continuous PNAD. It is of interest that the variables of the two samples, PSID and PNAD, be similar. Thus, the 2015 PNAD data will be used only to illustrate the proposed methodology.}. 
The PNAD data used in this work consists of a sample composed of women aged between 18 and 65 years old, with information on hourly wages and socio-economic characteristics. In total, the sample contains 26,460 observations, of which 387 are censored with a salary equal to zero. The data covers ten metropolitan regions in Brazil (Belém-PA, Fortaleza-CE, Recife-PE, Salvador-BA, Belo Horizonte- MG, Rio de Janeiro-RJ, Curitiba-PR, Porto Alegre-RS, Brasília-DF and São Paulo-SP). Nominal income values were deflated by the National Consumer Price Index (INPC) provided by the IBGE.

The objective is to study women's labor supply. The dependent variable is women's hourly wages ($T$) and the explanatory variables are: $age$, woman's age; $age^2$, woman's age squared; $color$, dummy for color with value 1 (white) or 0 (non-white); $civil$, dummy for marital status with value 1 (married) or 0 (non-married); $minor$, dummy for children under 10 in the household with a value of 1 (yes) or 0 (no); $educ$, captures educational returns on income and is classified by formal years of study ranging from 0 to 16 years; $exper$, captures the returns of the number of years in the main job on income, which is classified by years of work and can vary between 0 to 56 years; $head$, is a dummy that captures the condition of the woman in the household, presenting a value of 1 if the woman in the household is head
of the family and 0 otherwise (non-head). Similarly to the first application, the dependent variable, woman's hourly wage, is added to one ($T+1$), such that $\Psi=1$.

The choice of the above-described explanatory variables is due to their importance in the female labor supply literature, in addition to the interest in similarity with the PSID data. The variable $educ$, for example, directly affects female participation rates in the labor market. On the other hand, the presence of women in the productive world does not depend only on market demand, there are other factors that can limit this participation, such as the presence or absence of minors in the household.

\begin{figure}[!ht]
\centering
\subfigure[]{\includegraphics[height=7cm,width=7cm]{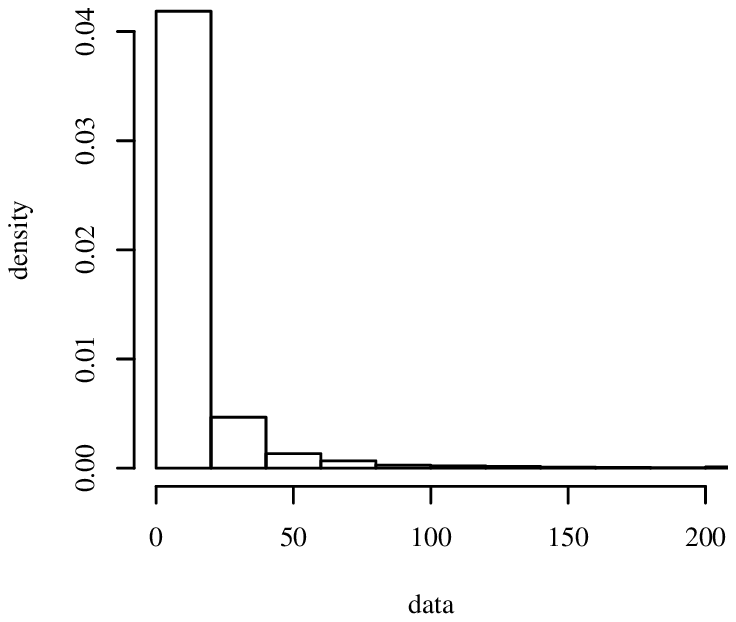}}
\subfigure[]{\includegraphics[height=6cm,width=6cm]{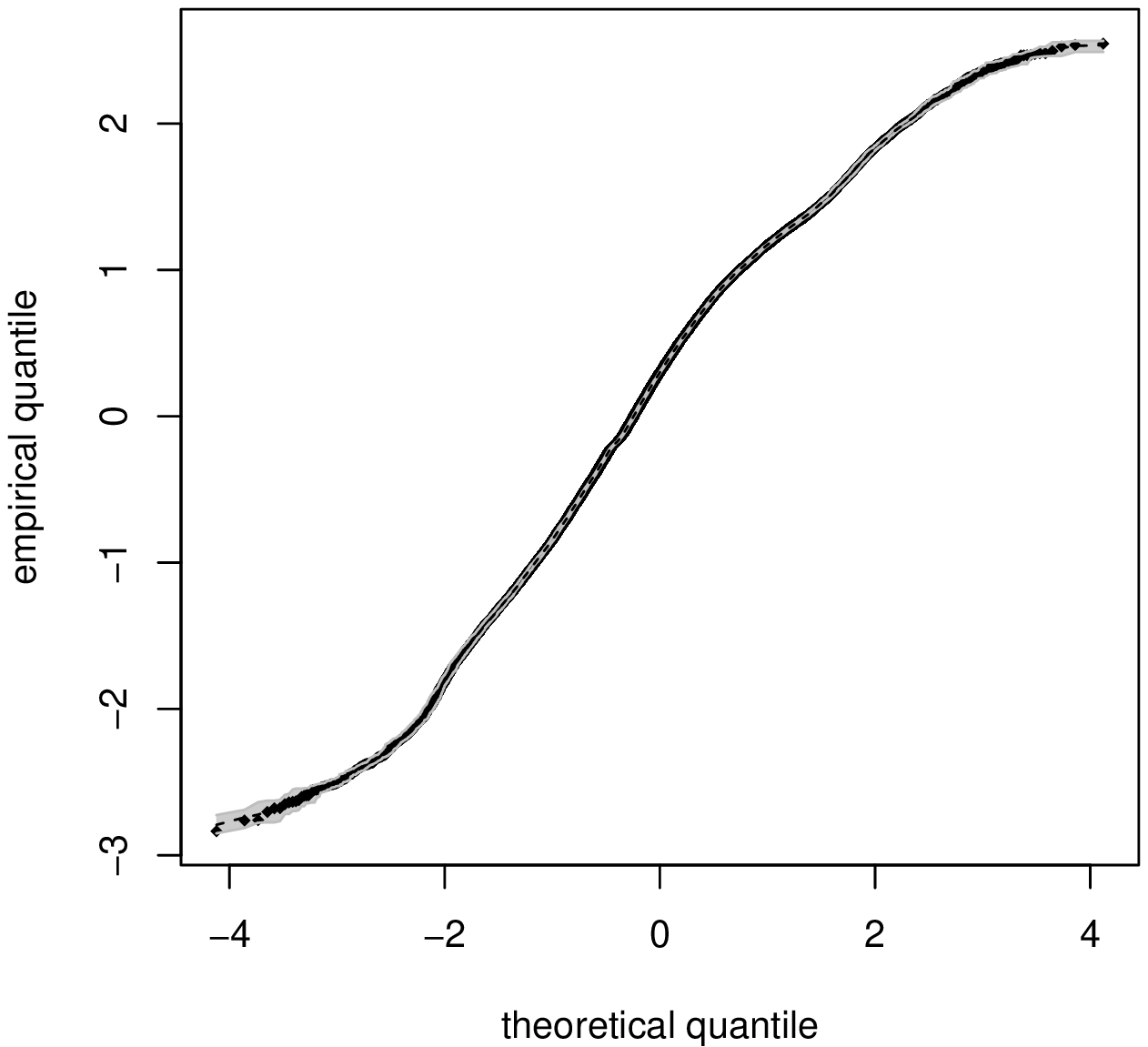}}
%\vspace{-0.25cm}
 \caption{\small {Histogram (a) for the observed women's hourly wages and QQ plot (b) and its envelope for the MT residual for the log-$t$ quantile tobit model ($q=0.50$).}}
\label{fig:histqqplot2}
\end{figure}

Descriptive statistics for women's hourly wages ($T$) indicate that the mean and median are 22.133 and 7.59, respectively. The coefficient of variation is 493.11\%, indicating a high dispersion of the data around the mean, whereas the coefficients of skewness and kurtosis are equal to 20.225 and 573.536, respectively. The result of the coefficient of skewness shows the presence of a high positive skewness and the coefficient of kurtosis indicates the presence of heavy tails, confirming the hypothesis of using log-symmetric distributions is plausible. The asymmetric nature of the data is confirmed by the histogram shown in Figure \ref{fig:histqqplot2}(a).

Similarly to the first application, two versions of the log-symmetric quantile tobit model are considered:
\begin{eqnarray*}
T_{i}=
\begin{cases}
\Psi, & \quad T_{i}^{\ast}  \leq  \Psi, \quad i=1,\ldots,387,\\
T_{i}^{\ast}=Q_i\epsilon_{i}^{\sqrt{\phi_i}}, & \quad  T_{i}^{\ast}  >  \Psi, \quad i=388,\ldots,26,460,
\end{cases}
\end{eqnarray*}
where 
$$Q_i=\exp\left( \beta_0 + \beta_1\, age_i + \beta_2\, age^2_i + \beta_3\, color_i  + \beta_4\, civil_i  + \beta_5\, minor_i\right.$$
 $$\left. +  \beta_6\, educ_i  +  \beta_7\, exper_i  +  \beta_8\, head_i\right),$$
$\epsilon_i \sim \textrm{QLS}(1, 1, g)$, where
\begin{itemize}
\item \textbf{Model 1}: $\phi_i=\exp\left(\kappa_0\right)$.
\item \textbf{Model 2}: $\phi_i=\exp\left( \kappa_0 + \kappa_1\, age_i + \kappa_2\, age^2_i + \kappa_3\, color_i  + \kappa_4\, civil_i  + \kappa_5\, minor_i\right.$

 $\left. +  \kappa_6\, educ_i  +  \kappa_7\, exper_i  +  \kappa_8\, head_i\right)$.
\end{itemize}
In Model 2, explanatory variables are present in the dispersion parameter $\phi$, that is, the presence of heteroscedasticity.

The AIC and BIC values for the adjusted log-symmetric quantile tobit models are presented in Table \ref{tab:resultadosaicbic:2}. In the case of the normal classic tobit model, the values are $\text{AIC}=313909.3$ and $\text{BIC}=313991.1$, which shows the superiority of the proposed models in terms of adjustment. In Table \ref{tab:resultadosaicbic:2}, the reported values of $ q $ are $ 0.05, 0.25, 0.50, 0.75 $ and $ 0.95 $, however, similar results are obtained considering $ q = \{0.01, 0.02, \ldots, 0.99\}$. It is noted that the models with explanatory variables in the dispersion parameter ($\phi$) present better adjustments in all the distributions considered. In general, the log-$t$ quantile tobit model presents the best fit, a result confirmed by the QQ plot with simulated envelope of the MT residual for this model with $ q = 0.50 $; see 
Figure \ref{fig:histqqplot2}(b). It is emphasized that similar plots are obtained for other values of $q$.

% latex table generated in R 3.6.3 by xtable 1.8-4 package
% Sun Jun  7 18:10:05 2020
\begin{table}[ht]
\footnotesize
\centering
 \caption{\small {AIC and BIC values for different models and $q$ with the PNAD data.}}
\begin{tabular}{lrrrrrrrrrrrrrr}
  \hline
           &Model     &  Criterion  &$q=0.05$  & $q=0.25$ & $q=0.50$ & $q=0.75$ & $q=0.95$  \\ 
					\hline
Log-NO     &  1        &  AIC       & 63738.61 & 63738.61 & 63738.61 & 63738.61 & 63738.61  \\        
           &  1        &  BIC       & 63820.45 & 63820.45 & 63820.45 & 63820.45 & 63820.45  \\  
           &  2        &  AIC       & 62632.53 & 62707.62 & 62748.85 & 62782.9  & 62822.39  \\       
           &  2        &  BIC       & 62779.83 & 62854.92 & 62896.15 & 62930.2  & 62969.69  \\[0.15cm] 
Log-$t$    &  1        &  AIC       & 53825.67 & 53825.67 & 53836.37 & 53825.67 & 53825.67 \\  
           &  1        &  BIC       & 53907.51 & 53907.51 & 53918.2  & 53907.51 & 53907.51 \\
           &  2        &  AIC       & 50348.87 & 51912.48 & 52303.72 & 52496.98 & 52696.98  \\       
           &  2        &  BIC       & 50496.17 & 52059.78 & 52451.02 & 52644.28 & 52844.28  \\ [0.15cm]
Log-PE     &  1        &  AIC       & 55905.30 & 55905.30 & 55905.30 & 55905.29 & 55905.30  \\        
           &  1        &  BIC       & 55987.13 & 55987.13 & 55987.13 & 55987.13 & 55987.14  \\  
           &  2        &  AIC       & 53986.88 &54591.75  & 54747.16 &54847.73  &54970.37  \\       
           &  2        &  BIC       & 54134.18 &54739.05  &54894.46  &54995.03  & 55117.67  \\[0.15cm] 
EBS        &  1        &  AIC       & 63895.42 &63895.42  &63895.42  &63895.42  &63895.42  \\        
           &  1        &  BIC       & 63977.25 &63977.25  &63977.25  &63977.25  &63977.25  \\  
           &  2        &  AIC       & 62806.21 & 62877.53 & 62917    &62949.78  & 62987.94 \\  
           &  2        &  BIC       & 62953.51 & 63024.83 & 63064.3  & 63097.08 & 63135.24 \\						
 \hline
\end{tabular}
\label{tab:resultadosaicbic:2}
\end{table}

The model parameter estimates for the log-$t$ tobit quantile model based on Model 2 considering $ q =\{ 0.05, 0.25, 0.50, 0.75, 0.95\} $ and $ q_{otm}$, are presented in Table \ref{tab:resultadosquantis:logt}. Note that the maximum likelihood estimates of the model parameters change according to the value of $q$, that is, the magnitude of the effect of the explanatory variables varies with $q$. Again, we can interpret the estimated coefficients in terms of the effect on the latent variable $T_i^*$ (observed part of the hourly wage). It is observed, for example, that for white women there is an increase in the $5^{\circ}$ percentile ($q=0.05$) of the hourly wage of $(\exp(0.0219)-1)*100\%=2.21\%$ when compared to non-whites. However, the increase in the $95^{\circ}$ percentile ($q=0.95$) of the hourly wage is of $(\exp(0.0219)-1)*100\%=26.05\%$. That is, the effect of $color$ on the observed part of the hourly wage is greater for women with higher income (larger quantiles). It is also observed in Table \ref{tab:resultadosquantis:logt} that the parameters associated with the explanatory variables $age$, $age^2$, $color$, $educ$, $exper$ and $head$ that model the dispersion, are significant, indicating the presence of heteroscedasticity in the data.

% latex table generated in R 3.6.3 by xtable 1.8-4 package
% Sun Jun  7 18:10:05 2020
\begin{table}[!ht]
\footnotesize
\centering
 \caption{\small {Maximum likelihood estimates (standard errors in parentheses) for the log-$t$ quantile tobit model across different values of $q$ (Model 2).}}
\begin{tabular}{lrrrrrrrrrrrrrr}
  \hline
                           & $q=0.05$ & $q=0.25$ & $q=0.50$ & $q=0.75$ & $q=0.95$  & $q_{otm}=0.01$\\ 
  \hline
$\beta_0\,(intercept)$     & 1.7685*    & 0.9726*   & 0.6617*   & 0.5392*   & 0.5919*  & 1.9216*\\ 
                           & (0.0522)   & (0.0351)  & (0.0304)  & (0.0367)  & (0.0731) & (0.0865)\\  
$\beta_1\,(age)$           &-0.0113*    & 0.0195*   & 0.0311*   & 0.0417*   &  0.0656* & -0.0413*\\  
                           & (0.0027)   & (0.0018)  & (0.0016)  & (0.0020)  & (0.0041) & (0.0044)\\ 
$\beta_2\,(age^2)$         & 0.0001*    & -0.0002*  & -0.0003*  & -0.0004*  & -0.0006* & 0.0004*\\  
                           & (0.0000)   & (0.0000)  & (0.0000)  & (0.0000)  & (0.0001) & (0.0001)\\  
$\beta_3\,(color)$         & 0.0219*    & 0.1235*   & 0.1683*   & 0.2003*   & 0.2315*  & -0.0735*\\ 
                           & (0.0108)   & (0.0068)  & (0.0062)  & (0.0081)  & (0.0163) & (0.0169)\\
$\beta_4\,(civil)$         &-0.0008     & -0.0177   & -0.0104   &  0.0038   &  0.0351  & 0.0399\\  
                           & (0.0236)   & (0.0167)  & (0.0161)  & (0.0207)  & (0.0412) & (0.0348)\\
$\beta_5\,(minor)$         & -0.0344*   & -0.0238*  & -0.0162*  & -0.0110   & -0.0020  & -0.0335**\\  
                           & (0.0111)   & (0.0070)  & (0.0064)  & (0.0082)  & (0.0162) & (0.0174)\\
$\beta_6\,(educ)$          & -0.0376*   &  0.0382*  & 0.0646*   & 0.0767*   &  0.0842* & -0.1227*\\  
                           & (0.0014)   & (0.0014)  & (0.0011)  & (0.0011)  & (0.0018) & (0.0019)\\
$\beta_7\,(exper)$         & -0.0012    & 0.0109*   & 0.0174*   & 0.0229*   & 0.0369*  & -0.0101*\\  
                           & (0.0009)   & (0.0007)  & (0.0006)  & (0.0008)  & (0.0017) & (0.0012)\\												
$\beta_8\,(head)$          & -0.0459*   & -0.0044   & 0.0197*   & 0.0397*   & 0.0798*  & -0.0811* \\ 
                           & (0.0110)   & (0.0075)  & (0.0069)  & (0.0087)  & (0.0175) & (0.0163)\\[0.30cm]
$\kappa_0\,(intercept)$    & -6.3627*   & -5.9789*   & -5.2843*   & -4.8397*   & -4.2542* & -5.2983*\\  
                           & (0.1141)   & (0.1608)   & (0.1559)   & (0.1466)   & (0.1370) & (0.0792)\\  
$\kappa_1\,(age)$          & 0.0829*    &  0.0930*   & 0.0916*    & 0.0884*    & 0.0806*  & 0.0579*\\  
                           & (0.0058)   & (0.0079)   & (0.0079)   & (0.0075)   & (0.0070) & (0.0040)\\  
$\kappa_2\,(age^2)$        & -0.0008*   & -0.0008*   & -0.0008*   & -0.0008*   & -0.0007* & -0.0006*\\  
                           & (0.0001)   & (0.0001)   & (0.0001)   & (0.0001)   & (0.0001) & (0.0000)\\  
$\kappa_3\,(color)$        & 0.2495*    & 0.2688*    & 0.2368*    & 0.2005*    & 0.1130   & 0.1799*\\  
                           & (0.0207)   & (0.0272)   & (0.0280)   & (0.0279)   & (0.0260) & (0.0142)\\ 
$\kappa_4\,(civil)$        & -0.0311    &  0.0649    &  0.0927    &  0.0990    & -0.0023  & -0.0572**\\  
                           & (0.0504)   & (0.0664)   & (0.0666)   & (0.0649)   & (0.0600) & (0.0328)\\ 
$\kappa_5\,(minor)$        & 0.0370     & 0.0314     & 0.0263     & 0.0251     & 0.0849   & 0.0194\\ 
                           & (0.0226)   & (0.0298)   & (0.0302)   & (0.0298)   & (0.0271) & (0.0152)\\
$\kappa_6\,(educ)$         &  0.2022*   & 0.1381*    & 0.0830*    & 0.0529*    & 0.0234*  & 0.1673*\\  
                           & (0.0030)   & (0.0048)   & (0.0042)   & (0.0036)   & (0.0028) & (0.0019)\\  
$\kappa_7\,(exper)$        & 0.0242*    & 0.0271*    & 0.0253*    & 0.0245*    & 0.0259*  & 0.0174*\\  
                           & (0.0014)   & (0.0019)   & (0.0020)   & (0.0021)   & (0.0020) & (0.0009)\\ 
$\kappa_8\,(head)$         & 0.1104*    & 0.1227*    & 0.1142*    & 0.1070*    & 0.0938*  & 0.0795*\\  
                           & (0.0211)   & (0.0289)   & (0.0292)   & (0.0288)   & (0.0265) & (0.0139)\\ 													
$\xi$                      & 2 & 2 & 2 & 2 & 2 & 2\\   
    \hline \multicolumn{6}{l}{\scriptsize{Source: Elaborated by the authors based on PNAD data.}}\\[-0.1cm]
        \multicolumn{6}{l}{\scriptsize{* significant at 5\% level. ** significant at 10\% level.}}
\end{tabular}
\label{tab:resultadosquantis:logt}
\end{table}

\section{Concluding remarks}\label{sec:6}
In this work, a class of quantile tobit models was proposed based on a reparameterization of the log-symmetric distributions. In such reparameterization, the quantile is one of the distribution parameters. The advantages of the proposed models over the classic tobit model include: 
\begin{itemize}
 \item[(i)] flexibility to assume several asymmetric distributions, since the quantile-based log-symmetric class incorporates several distributions as special cases, such as log-normal, log-Student-$t$, log-power-exponential and extended Birnbaum-Saunders, among others;
 \item[(ii)] greater flexibility in the analysis of the effects of explanatory variables on the dependent variable due to the quantile approach; and
 \item[(iii)] ability to accommodate heteroscedasticity, since the proposed model allows the insertion of explanatory variables in the dispersion parameter.
\end{itemize}

A Monte Carlo simulation study was carried out to evaluate the performance of the maximum likelihood estimates. In general, the results have shown good performances of the maximum likelihood estimates in terms of bias and mean squared error. Two applications to real data from PSID and PNAD were carried out to illustrate the proposed methodology. The applications have favored the use of the proposed log-symmetric quantile tobit models over the classic tobit model, illustrating the advantages (i), (ii) and (iii). As the final product of this paper, the authors are preparing an \texttt{R} package \citep{r2020vienna}, which can be an important tool for many professionals, researchers in the field of economics and statistics, data scientists, among others.

For future research, the following lines can be explored:
 \begin{itemize}
  \item[(i)] to study some hypothesis and misspecification tests via Monte Carlo simulation; see \cite{santoscribari:17}; 
   \item[(ii)] to generalize the proposed models for the cases with right censoring or two-sided censoring; see \cite{long:97}[pp. 211-212]; and
  \item[(iii)] to propose bivariate models; see \cite{yoo2005}.
  \item[(iv)] to propose zero-adjusted log-symmetric quantile regression models; see \cite{Heller2006}.
 \end{itemize}
Work on these problems is currently in progress and we hope to report these findings in future papers.

\normalsize

% % % \bibliographystyle{apalike}
% % % \bibliography{referencias}

\begin{thebibliography}{}

\bibitem[Alhamzawi and Ali, 2018]{alhamzawiali:18}
Alhamzawi, R. and Ali, H. T.~M. (2018).
\newblock Bayesian tobit quantile regression with penalty.
\newblock {\em Communications in Statistics - Simulation and Computation},
  47(6):1739--1750.

\bibitem[Amemiya, 1984]{amemiya:84}
Amemiya, T. (1984).
\newblock Tobit models: A survey.
\newblock {\em Journal of Econometrics}, 24:3--61.

\bibitem[Barros et~al., 2018]{bgls:17}
Barros, M., Galea, M., Leiva, V., and Santos-Neto, M. (2018).
\newblock {Generalized tobit models: Diagnostics and application in
  econometrics}.
\newblock {\em Journal of Applied Statistics}, 45:145--167.

\bibitem[Barros et~al., 1995]{barrosjatobamendoca95}
Barros, R., Jatob\'a, J., and Mendon\c{c}a, R. (1995).
\newblock A evolução da participa\c{c}\~ao de mulheres no mercado de
  trabalho: uma an\'alise de decomposi\c{c}\~ao.
\newblock {P}roceedings of the 4th {N}ational {M}eeting of {L}abor {S}tudies,
  Brazilian Association of Labor Studies.

\bibitem[Buchinsky, 1998]{buchinsky:98}
Buchinsky, Moshe;~Hahn, J. (1998).
\newblock An alternative estimator for the censored quantile regression model.
\newblock {\em Econometrica}, 66:653--671.

\bibitem[Cox and Hinkley, 1974]{ch:74}
Cox, D. and Hinkley, D. (1974).
\newblock {\em {Theoretical Statistics}}.
\newblock Chapman and Hall, London, UK.

\bibitem[Davino et~al., 2014]{dfv:14}
Davino, C., Furno, M., and Vistocco, D. (2014).
\newblock {\em Quantile Regression}.
\newblock Wiley, Chichester, UK.

\bibitem[Desousa et~al., 2018]{desousaetal:18}
Desousa, M.~F., Saulo, H., Leiva, V., and Scalco, P. (2018).
\newblock On a tobit-{B}irnbaum-{S}aunders model with an application to medical
  data.
\newblock {\em Journal of Applied Statistics}, 45:932--955.

\bibitem[Fair, 1977]{fair:77}
Fair, R. (1977).
\newblock A note on computation of the {T}obit estimator.
\newblock {\em Econometrica}, 45:1723--1727.

\bibitem[Fair, 1978]{fair:78}
Fair, R. (1978).
\newblock A theory of extramarital affairs.
\newblock {\em Journal of Political Economy}, 86:45--61.

\bibitem[Greene, 2012]{Greene2003Econometric}
Greene, W.~H. (2012).
\newblock {\em Econometric Analysis}.
\newblock Pearson Education, seventh edition.

\bibitem[Hao and Naiman, 2007]{hn:07}
Hao, L. and Naiman, D. (2007).
\newblock {\em Quantile Regression}.
\newblock Sage Publications, California , US.

\bibitem[Heckman and MaCurdy, 1980]{heckmanmaCurdy:80}
Heckman, J.~J. and MaCurdy, T.~E. (1980).
\newblock A life cycle model of female labor supply.
\newblock {\em Review of Economic Studies}, 47:47--74.

\bibitem[Heller et~al., 2006]{Heller2006}
Heller, G., Stasinopoulos, M., and Rigby, B. (2006).
\newblock The zero-adjusted inverse gaussian distribution as a model for
  insurance claims.
\newblock In J.~Hinde, J.~E. and Newell, J., editors, {\em Proceedings of the
  21th International Workshop on Statistical Modelling}, pages 226--233,
  Galway, Ireland. Statistical Modelling Society, University of Lancaster.
  
\bibitem[Helsel, 2011]{helsel:11}
Helsel, D.~R. (2011).
\newblock {\em Statistics for Censored Environmental Data Using Minitab and R}.
\newblock John Wiley \& Sons, Hoboken, New Jersey.

\bibitem[ILO, 2018]{iol2018}
ILO (2018).
\newblock {\em World Employment and Social Outlook: Trends for Women 2018 –
  Global snapshot}.
\newblock International Labour Organization, Geneva.

\bibitem[Islam, 2007]{islam:07}
Islam, N. (2007).
\newblock A dynamic tobit model of female labor supply.
\newblock {\em Working Papers In Economics No 259}, pages 1--29.

\bibitem[Jacobsen, 1999]{jacobsen99}
Jacobsen, J. (1999).
\newblock Labor force participation.
\newblock {\em The Quarterly Review of Economics and Finance}, 39(5):597--610.

\bibitem[Jarque, 1987]{jarque:87}
Jarque, C. (1987).
\newblock An application of {LDV} models to household expenditure analysis in
  {M}exico.
\newblock {\em Journal of Econometrics}, 36:31--54.

\bibitem[Ji et~al., 2012]{jilinzhang:12}
Ji, Y., Lin, N., and Zhang, B. (2012).
\newblock Model selection in binary and tobit quantile regression using the
  gibbs sampler.
\newblock {\em Computational Statistics \& Data Analysis}, 56(4):827 -- 839.

\bibitem[Jones, 2008]{j:08}
Jones, M.~C. (2008).
\newblock On reciprocal symmetry.
\newblock {\em Journal of Statistical Planning and Inference}, 138:3039--3043.

\bibitem[Kano et~al., 1993]{kanoetal93}
Kano, Y., Berkane, M., and Bentler, P.~M. (1993).
\newblock Statistical inference based on pseudo-maximum likelihood estimators
  in elliptical populations.
\newblock {\em Journal of the American Statistical Association},
  88(421):135--143.

\bibitem[Koenker, 2005]{koenker:05}
Koenker, R. (2005).
\newblock {\em Quantile Regression}.
\newblock Cambridge University Press, Cambridge.

\bibitem[Koenker and {Bassett Jr}, 1978]{koenker:78}
Koenker, R. and {Bassett Jr}, G. (1978).
\newblock Regression quantiles.
\newblock {\em Econometrica}, 46:33--50.

\bibitem[Long, 1997]{long:97}
Long, J.~S. (1997).
\newblock {\em Regression Models for Categorical and Limited Dependent
  Variables}.
\newblock Sage Publications, Inc., Thousand Oaks, US.

\bibitem[Lucas, 1997]{l:97}
Lucas, A. (1997).
\newblock {Robustness of the student $t$ based M-estimator}.
\newblock {\em Communications in Statistics: Theory and Methods},
  41:1165--1182.

\bibitem[Medeiros and Ferrari, 2017]{medeirosferrari:16}
Medeiros, M.~C. and Ferrari, S. L.~P. (2017).
\newblock Small-sample testing inference in symmetric and log-symmetric linear
  regression models.
\newblock {\em Statistica Neerlandica}, 71:200--224.

\bibitem[Melenberg and van Soest, 1996]{melenbergsoest:96}
Melenberg, B. and van Soest, A. (1996).
\newblock Parametric and semi-parametric modelling of vacation expenditures.
\newblock {\em Journal of Applied Econometrics}, 11:59--76.

\bibitem[Moffitt, 1982]{moffitt:82}
Moffitt, R. (1982).
\newblock The tobit model, hours of work and institutional constraints.
\newblock {\em The Review of Economics and Statistics}, 64:510--515.

\bibitem[Mroz, 1987]{m:87}
Mroz, T. (1987).
\newblock The sensitiviy of an empirical model of married women's hours of work
  to economic and statistical assumptions.
\newblock {\em Econometrica}, 55:765--799.

\bibitem[Powell, 1986]{powel:86}
Powell, J.~L. (1986).
\newblock Censored regression quantiles.
\newblock {\em Journal of Econometrics}, 32:143--155.

\bibitem[{R Core Team}, 2020]{r2020vienna}
{R Core Team} (2020).
\newblock {\em R: A Language and Environment for Statistical Computing}.
\newblock R Foundation for Statistical Computing, Vienna, Austria.

\bibitem[Santos and Cribari-Neto, 2017]{santoscribari:17}
Santos, J. and Cribari-Neto, F. (2017).
\newblock Hypothesis testing in log-{B}irnbaum-{S}aunders regressions.
\newblock {\em Communications in Statistics - Simulation and Computation},
  46:3990--4003.

\bibitem[Saulo et~al., 2020a]{ssls:20}
Saulo, H., Dasilva, A., Leiva, V., and S\'anchez, L. (2020a).
\newblock Log-symmetric quantile regression models.
\newblock {\em arXiv available at https://arxiv.org/abs/2010.09176}, pages
  1--31.

\bibitem[Saulo et~al., 2020b]{slnb:20}
Saulo, H., Le\~ao, J., Nobre, L., and Balakrishnan, N. (2020b).
\newblock {A class of asymmetric regression models for left-censored data}.
\newblock {\em Brazilian Journal of Probability and Statistics}, pages 1--23.

\bibitem[Scorzafave and Menezes-Filho, 2001]{scomenfilho01}
Scorzafave, L.~G. and Menezes-Filho, N.~A. (2001).
\newblock Participação feminina no mercado de trabalho brasileiro: evolução
  e determinantes.
\newblock {\em Pesquisa e Planejamento Econômico}, 31:441--478.

\bibitem[Seung-Hoon, 2005]{yoo2005}
Seung-Hoon, Y. (2005).
\newblock Analysing household bottled water and water purifier expenditures:
  simultaneous equation bivariate tobit model.
\newblock {\em Applied Economics Letters}, 12:297--301.

\bibitem[Silva et~al., 2009]{SILVA20094482}
Silva, G.~O., Ortega, E.~M., and Cordeiro, G.~M. (2009).
\newblock A log-extended weibull regression model.
\newblock {\em Computational Statistics \& Data Analysis}, 53(12):4482 -- 4489.

\bibitem[Stute, 1992]{W-1992}
Stute, W. (1992).
\newblock Strong consistency of the mle under random censoring.
\newblock {\em Metrika}, 39.

\bibitem[Therneau et~al., 1990]{tgf:90}
Therneau, T., Grambsch, P., and Fleming, T. (1990).
\newblock {Martingale-based residuals for survival models}.
\newblock {\em Biometrika}, 77:147--160.

\bibitem[Tobin, 1958]{t:58}
Tobin, J. (1958).
\newblock Estimation of relationships for limited dependent variables.
\newblock {\em Econometrica}, 26:24--36.

\bibitem[Vanegas and Paula, 2015]{vanegasp:15}
Vanegas, L.~H. and Paula, G.~A. (2015).
\newblock A semiparametric approach for joint modeling of median and skewness.
\newblock {\em Test}, 24:110--135.

\bibitem[Vanegas and Paula, 2016a]{vanegasp:16a}
Vanegas, L.~H. and Paula, G.~A. (2016a).
\newblock Log-symmetric distributions: statistical properties and parameter
  estimation.
\newblock {\em Brazilian Journal of Probability and Statistics}, 30:196--220.

\bibitem[Vanegas and Paula, 2016b]{vanegasp:16b}
Vanegas, L.~H. and Paula, G.~A. (2016b).
\newblock {\em ssym: Fitting Semi-Parametric log-Symmetric Regression Models}.
\newblock R package version 1.5.7.

\bibitem[Vanegas and Paula, 2017]{vanegaspaula:17}
Vanegas, L.~H. and Paula, G.~A. (2017).
\newblock Log-symmetric regression models under the presence of non-informative
  left- or right-censored observations.
\newblock {\em Test}, 26:405--428.

\bibitem[Weisberg, 2014]{weisberg:14}
Weisberg, S. (2014).
\newblock {\em Applied Linear Regression}.
\newblock John Wiley \& Sons, Hoboken, New Jersey, fourth edition.

\bibitem[Yue and Hong, 2012]{yuehong:12}
Yue, Y.~R. and Hong, H.~G. (2012).
\newblock Bayesian tobit quantile regression model for medical expenditure
  panel survey data.
\newblock {\em Statistical Modelling}, 12(4):323--346.

\end{thebibliography}

\end{document}